\documentclass[]{elsarticle}

\usepackage[T1]{fontenc}

\usepackage[frenchb,english]{babel}
\usepackage{amsmath}
\usepackage{amssymb}
\usepackage{graphicx}
\usepackage{dcolumn}
\usepackage{helvet}
\usepackage{courier}
\usepackage{bm}
\usepackage{microtype} 			

\numberwithin{equation}{section} 	
\let\originalleft\left						
\let\originalright\right					%
\renewcommand{\left}{\mathopen{}\mathclose\bgroup\originalleft}	%
\renewcommand{\right}{\aftergroup\egroup\originalright} 	%

\newcommand{\di}{\mathrm{d}}
\newcommand{\eqspace}{\,}
\newcommand{\omegalzero}{\omega_l^0}
\newcommand{\omegajzero}{\omega_j^0}
\newcommand{\xlzero}{{\bm x}_l^0}
\newcommand{\xjzero}{{\bm x}_j^0}
\renewcommand{\Pr}{\mathrm{Pr}}
\newcommand{\Ra}{\mathrm{Ra}}
\newcommand{\Nu}{\mathrm{Nu}}
\newcommand{\abbrev}{\star}

\begin{document}

\begin{frontmatter}

\selectlanguage{english}

\title{The Lundgren-Monin-Novikov Hierarchy:\\Kinetic Equations for Turbulence}

\author{R. Friedrich$\dagger^{1,2}$}
\author{A. Daitche$^1$}
\author{O. Kamps$^{2}$}
\author{J. L{\"u}lff$^1$}
\author{M. Vo{\ss}kuhle$^3$}
\author{M. Wilczek$^1$}
\ead{mwilczek@uni-muenster.de}

\address{$^1$ Institute for Theoretical Physics, Westf{\"a}lische Wilhelms-Universit{\"a}t, Wilhelm-Klemm-Str. 9, D-48149 M{\"u}nster, Germany}
\address{$^2$ Center for Nonlinear Science, Westf{\"a}lische Wilhelms-Universit{\"a}t, Corrensstr. 2, D-48149 M{\"u}nster, Germany}
\address{$^3$ Laboratoire de Physique, ENS de Lyon, 46 all\'{e}e d'Italie F-69007 Lyon, France}

\enlargethispage*{\baselineskip} 

\begin{abstract}
We present an overview of recent works on the statistical description of turbulent flows in terms of probability density functions (PDFs) in the framework of the Lundgren-Monin-Novikov (LMN) hierarchy.
Within this framework, evolution equations for the PDFs are derived from the basic equations of fluid motion.
The closure problem arises either in terms of a coupling to multi-point PDFs or in terms of conditional averages entering the evolution equations as unknown functions.
We mainly focus on the latter case and use data from direct numerical simulations (DNS) to specify the unclosed terms.
Apart from giving an introduction into the basic analytical techniques, applications to two-dimensional vorticity statistics, to the single-point velocity and vorticity statistics of three-dimensional turbulence, to the temperature statistics of Rayleigh-B\'{e}nard convection and to Burgers turbulence are discussed.

\selectlanguage{frenchb}
\vspace*{0.5\baselineskip}
\noindent
\enlargethispage{1.00\baselineskip}
Nous pr\'{e}sentons un aper\c{c}u des travaux r\'{e}cents sur la description statistique des \'{e}coulements turbulents en terme de fonctions densit\'{e} de probabilit\'{e} (PDFs) dans le cadre de la hi\'{e}rarchie de Lundgren-Monin-Novikov (LMN).
Dans ce cadre, des \'{e}quations d'\'{e}volution pour les PDFs sont d\'{e}riv\'{e}es \`{a} partir des \'{e}quations fondamentales d\'{e}crivant la dynamique des fluides.
Le probl\`{e}me de fermeture se pose soit sous la forme d'une couplage aux PDFs multipoints, soit sous la forme de moyennes conditionnelles qui entrent dans les \'{e}quations d'\'{e}volution en tant que fonctions inconnues.
Nous nous concentrons principalement sur le dernier cas et utilisons les donn\'{e}es des simulations num\'{e}riques directes (DNS) pour calculer les termes non ferm\'{e}s.
Nous donnerons donc une introduction de base aux techniques analytiques.
Ensuite, nous pr\'{e}senterons quelques applications, en particulier \`{a} la statistique de la vorticit\'{e} en deux dimensions, aux statistiques de vitesse et de vorticit\'{e} en un point en trois dimensions, \`{a} la statistique de la temp\'{e}rature dans la convection de Rayleigh-B\'{e}nard et \`{a} la turbulence de Burgers.

\selectlanguage{english}

\end{abstract}

\end{frontmatter}

\selectlanguage{english}
\tableofcontents
\newpage

\section{Introduction}
The challenge in fundamental turbulence research is the prediction of characteristic flow features starting from the basic equations of fluid motion.
For flows in the turbulent regime predictive results are rather sparse and, despite of important applications in science and engineering, basic properties of turbulent flows remain poorly understood starting from first principles \cite{Tsinober2009book}.

The same applies to statistical hydrodynamics, which aims at a description of turbulent flows based on a statistical treatment of the basic hydrodynamic equations \cite{monin07book1,monin07book2}.
In particular, solving the statistical equations derived from the basic hydrodynamic equations has only been successful in but a few cases with Kolmogorov's four-fifth law being one of the most prominent examples \cite{pope00book}.
The main reason for this is that the Navier-Stokes equation, which describes the fluid motion mathematically, is both nonlinear and nonlocal.
As a consequence, statistical relations derived from this equation are in general unclosed, such that a statistical description without further ad hoc assumptions usually leads to an infinite set of statistical equations or, in a more condensed notation, to functional equations \cite{monin07book1,monin07book2}.
The investigation of these statistical equations and the development of physically reasonable closures may be seen as one of the central issues of theoretical turbulence research.

The statistical investigation of turbulent flows has a long history, with its origin in the works of Reynolds \cite{Reynolds1895}, who formulated moment equations for the mean and fluctuating velocity.
Already this pioneering work has revealed the fundamental problem of statistical turbulence research, the closure problem, showing up in terms of the turbulent Reynolds stresses.
The infinite hierarchy of moment equations was first formulated by Friedmann and Keller \cite{Keller1924pro} and has become the major line of research during the first half of the twentieth century.
The challenge consists in adding further exact relations to Kolmogorov's four-fifth law, which allows one to formulate exact or approximate closures for the hierarchy \cite{Yakhot2001prl,Falkovich2010jfm}.
The formulation of an ad hoc closure for a higher-order moment equation eventually may lead  to a realizability problem, e.\,g.~the generation of negative energy.
This realizability problem arises in the naive Gaussian closure formulated by Millionshchikov \cite{Millionshchikov1941dan}.

Instead of formulating ad hoc closures of the Friedmann-Keller hierarchy, it is possible to study the statistics in the framework of the Lundgren-Monin-Novikov hierarchy, which often also is referred to as  ``PDF methods'' or ``kinetic equations''.
Work on this has been initiated in the sixties of the last century by Lundgren, Monin and Novikov \cite{lundgren67pof,monin67pmm,novikov68sdp}.
In this theoretical framework, evolution equations for the probability density functions of turbulent observables are obtained from the equations of motion of the fluid.
Generally speaking, kinetic equations are conservation laws for the probability distributions of the vorticity or velocity field at a number of spatial points and time instants.
They are obtained by relating the dynamics of the $N$-point probability distributions to the dynamics of the underlying classical field theory.
The result is an infinite hierarchy of evolution equations for the $N$-point distribution functions of the form
\begin{equation}\label{LMNshort}
  \frac{\partial }{\partial t}f_N=\mathrm{L}_N f_N+\mathrm{L}_{N,N+1}f_{N+1} \eqspace .
\end{equation}
Here, $f_N$ and $f_{N+1}$ are the $N$ and $(N+1)$-point PDFs, which are related by operators $\mathrm{L}_N$ and $\mathrm{L}_{N,N+1}$, where the latter explicitly highlights the
coupling to the higher-order probability density.
While Lundgren and Monin focused on the velocity statistics, Novikov studied the statistics of the vorticity field.
Of course, unclosed terms arise also in this framework, which can be expressed either in terms of the statistics of an increased number of points in space or in terms of conditional averages, which then enter the PDF equations as unknown functions.
Lundgren and Monin mainly followed the former way and investigated the arising hierarchy with respect to possible analytical closures.
Additionally Lundgren pointed out interesting analogies to the BBGKY hierarchy of kinetic gas theory \cite{bogolyubova66book,lifshitz81book}.
Novikov emphasized the possibility to introduce conditional expectations.
A particularly interesting paper is the one of Ulinich and Lyubimov \cite{ulinich69spj}, which shortly appeared after the works of Lundgren, Monin and Novikov.

It can be shown that the LMN approach is completely equivalent to the statistical description of turbulence by moment equations.
In fact, the Friedmann-Keller equations for the $N$-point-single-time correlation functions of the field can be obtained by deriving the evolution equations for the
expectations
\begin{equation}
  \langle \omega({\bm x}_1,t)\dots\omega({\bm x}_N,t) \rangle = \int \! \di \omega_1\dots \di \omega_N \, \omega_1\dots\omega_N \, f(\lbrace\omega_l,{\bm x}_l\rbrace,t)
\end{equation}
in a straightforward manner, here exemplified for the pseudo-scalar vorticity of two-dimensional turbulence.
The infinite set of evolution equations for the correlation functions, which frequently are also denoted as the Hopf equations, can be summarized in a condensed notion in the Hopf-functional equation for the characteristic functional
\begin{equation}
  Z[\alpha]=\left\langle \exp\left[\mathrm{i} \int \! \mathrm{d}{\bm x}\, \alpha({\bm x}) \omega({\bm x},t)\right] \right\rangle \eqspace .
\end{equation}
In this sense, the LMN hierarchy is equivalent to Hopf's functional equation.
In fact, Monin showed that the hierarchy of PDF equations corresponds to the Hopf functional equation formulated at discrete points in space.

The motivation for the study of the LMN hierarchy is similar to the one for the examination of the Friedmann-Keller equation.
The aim is to find a closure of the chain of equations for low values of $N$, which allows one to assess the statistics of the field restricted to $N$ points in space.
The advantage of formulating closures for the LMN hierarchy compared to the Friedmann-Keller hierarchy consists in the fact that the formulation of realizable closures, i.\,e.~closures which do not generate spurious results like negative energy spectra, is more transparent compared to closures for the Friedmann-Keller hierarchy.
This has been stressed essentially by Lundgren and, later on, has been advocated by Pope in his work on turbulent combustion \cite{pope85pec}.
Furthermore, there is a (formal) analogy to the kinetic theory of many-particle systems and the derivation of kinetic equations using the hierarchy of Bogoljubov, Born, Green, Kirkwood, Yvon (BBGKY hierarchy \cite{bogolyubova66book,lifshitz81book}), and it is tempting to apply decoupling procedures developed in the kinetic theory of gases and plasmas to the LMN hierarchy.

Besides the possibility to formulate realizable closures, the LMN hierarchy has another appealing property, which is related to progress in the field of direct numerical simulations (DNS).
This line of research actually has been initiated by Novikov who started his work with the infinite set of evolution equations
while later focusing on the conditional vorticity field combining analytical closures and numerical investigations of the field.
The introduction of conditional probability distributions 
\begin{equation}
  f_{N+1}=p_{N+1,N}f_N
\end{equation}
allows one to formulate (formally closed) kinetic equations for the $N$-point probability distribution: 
\begin{equation}
  \frac{\partial }{\partial t} f_N=[\mathrm{L}_N+\mathrm{M}_{N}]f_N \qquad , \qquad \mathrm{M}_N=\mathrm{L}_{N,N+1} p_{N+1,N}
\end{equation}
Progress in DNS allows one to assess the conditional PDFs for the low $N$ case, especially for $N=1,2$.
An interesting conclusion from Novikov's work is that the statistical equations describing the vorticity are incompatible with the assumption that the vorticity field can be represented by a Gaussian random field \cite{novikov68sdp}.
This shows that a statistical description of this non-equilibrium system has to rely on the usage of non-Gaussian random fields.
This is certainly one of the main conceptual challenges of modern turbulence research.

PDF equations in combination with conditional averages have also been established by Pope for the investigation of reactive flows and combustion, based
on the observation that the reaction kinetics is local in space, and, hence does not require the formulation of a closure
in the corresponding PDF equation.
Combined with stochastic modeling, a whole branch of research with engineering applications has emerged.
Although the starting point of Pope's investigations is very close to the work of Lundgren, his line of research aims at modeling and applications \cite{pope85pec,pope00book}, especially with respect to single-point closures.

Theoretical research on the LMN hierarchy remains an active area of research up to today.
Recent work by Yakhot deals with analytical closures and the application to scalar fields and convection \cite{sinai89prl,yakhot89prl,yakhot98pre}.
Work on the statistics of velocity fields has been pursued by Hosokawa as well as Tatsumi et al., with the aim of formulating analytical closures of the PDF equation \cite{hosokawa08pre,tatsumi04fdr,tatsumi07jfr}.
An especially interesting step here was taken by Tatsumi with his so-called cross-independence hypothesis, which neglects statistical correlations of the increments and sums of velocities at two points in space, yielding a non-trivial closure to the statistical equations \cite{tatsumi11jfm}.
Boffetta et al.~\cite{boffetta2002Bpre} used DNS to formulate a closure for the statistics of velocity increments of two-dimensional turbulence.
Ching \cite{ching93prl} considered applications to convection, and furthermore, together with Pope, formulated equations for the PDFs containing conditional averages \cite{pope1993pof,ching96pre}.
These relations are based on statistical symmetries, and are reviewed below.

Finally, we mention the work of Polyakov \cite{Polyakov1995pre} on the statistical formulation of the forced Burgers turbulence.
He formulated the evolution equation for the single-point PDF and a closure of the dissipative term.
His work stimulated a variety of theoretical investigations, summarized in the review of B\'{e}c and Khanin \cite{Bec2007pr}.

Whereas the approaches based on the formulation of kinetic equations aim at assessing the statistics of the turbulent fields starting from exact relations derived from the basic fluid dynamics equation, a completely different approach has been pursued in the works of Friedrich, Peinke and coworkers \cite{Friedrich1997prl,Lueck1999prl,Renner2001jfm,Renner2003prl, Friedrich2011pr}.
In this approach a phenomenological theory has been developed, which is based on a Fokker-Planck equation for the PDFs of the velocity increments.
The evolution of the velocity increments ${\bm v}(\bm r,t)={\bm u}({\bm x}+{\bm r},t)-{\bm u}({\bm x},t)$ is considered as a Langevin process in scale $r$.
Empirical investigations of this process prove Markovian properties on scales larger than the Einstein-Markov scale \cite{Lueck2006pla} (which is of the order of the Taylor length).
Furthermore, estimation of the corresponding Kramers-Moyal coefficients allows one to formulate a Fokker-Planck equation for the transition PDFs as a function of scale.
In this way one gets access to the $N$-scale velocity increment PDFs and obtains a phenomenological description of the phenomenon of intermittency.
One may hope that eventually the kinetic equation approach to turbulence based on the LMN hierarchy, which is the topic of the present review, may lead to a derivation of the phenomenological Fokker-Planck theory of intermittency of the turbulent cascade.

\subsection{Outline of the Review}
The aim of the present review is to summarize a number of recent works applying the statistical equations of the LMN hierarchy to some basic turbulent flows.
It is structured as follows.
In the next section we summarize basic concepts and methods of the kinetic equation approach to turbulent fields.
Then, we review recent work on the two-point statistics of two-dimensional turbulence, and single-point statistics of velocity and vorticity of fully developed three-dimensional turbulence.
As an example of a confined turbulent flow we discuss Rayleigh-B\'{e}nard convection with emphasis on the boundary layer statistics of the turbulent field.
We end up with some analytical considerations of Burgers turbulence before we conclude.

\section{Kinetic Equations: Concepts and Methods} \label{sec:concepts}
In this subsection we introduce the basic definitions and some relations that are necessary for the statistical description of classical random fields.
For the example of the vorticity statistics of two-dimensional turbulence we show how these techniques can be utilized for the  derivation of the evolution equations of the LMN hierarchy.
For a very readable account on this topic we refer the reader to the paper by Lundgren \cite{lundgren67pof} or the textbook by Pope \cite{pope00book}.

\subsection{Probability Distributions and Random Fields}
The starting point is the definition of the so-called  {\em fine-grained probability distribution} for the vorticity field at a base point ${\bm x}_1$ at time $t$,
\begin{equation}
  \hat f_1(\omega_1,{\bm x}_1,t) =\delta\bigl(\omega_1-\omega({\bm x}_1,t)\bigr) \eqspace .
\end{equation}
Here $\omega({\bm x}_1,t)$ denotes a realization of the vorticity field and $\omega_1$ is the corresponding sample space variable.
Therefore, $\hat{f}_1$ is a PDF with respect to $\omega_1$ and a function with respect to the variables ${\bm x}_1$ and $t$.
The fine-grained PDFs, loosely speaking, provide a probabilistic description of a single realization of the random field.
In order to obtain the full PDFs from the fine-grained PDFs, one has to perform an ensemble average over all possible realizations of the flow leading to
\begin{equation}
  f_1(\omega_1,{\bm x}_1,t) =  \langle \hat f_1(\omega_1,{\bm x}_1,t) \rangle = \bigl\langle \delta\bigl(\omega_1-\omega({\bm x}_1,t)\bigr)\bigr\rangle \eqspace ,
\end{equation}
where the brackets denote the averaging process.
Both definitions can be easily extended to $N$ points yielding
\begin{align}
  f_N(\omega_1,{\bm x}_1, \dots , \omega_N,{\bm x}_N,t) & = \langle \hat f_N(\omega_1,{\bm x}_1, \dots, \omega_N,{\bm x}_N,t) \rangle \nonumber \\ &= 
  \left \langle \prod_{i=1}^N \delta\bigl(\omega_i-\omega({\bm x}_i,t)\bigr) \right \rangle  \eqspace .
\end{align}

In general the PDFs have to fulfill some constraints like the {\em normalization} property
\begin{equation}
  \int \! \mathrm{d}\omega_1 \dots \int \! \mathrm{d}\omega_N \, f_N = 1
\end{equation}
and the {\em reduction} property 
\begin{equation}
  \int \! \mathrm{d}\omega_N \, f_N = f_{N-1} \eqspace .
\end{equation}
Due to the fact that we investigate physically meaningful fields, the PDFs are also subject to additional constraints like the separation property
\begin{equation}
  \lim_{|{\bm x}_2 -{\bm x}_1 |\rightarrow \infty} f_2(\omega_1,{\bm x}_1,\omega_2,{\bm x}_2,t)= f_1(\omega_1,{\bm x}_1,t)f_1(\omega_2,{\bm x}_2,t)
\end{equation}
which means the vorticity at two different points should become statistically independent when the distance between them is large enough, 
and the {\em coincidence} property
\begin{equation}\label{fusion}
  \lim_{|{\bm x}_2 -{\bm x}_1 |\rightarrow 0} f_2(\omega_1,{\bm x}_1,\omega_2,{\bm x}_1,t)=\delta\bigl(\omega_1-\omega_2\bigr) f_1(\omega_1,{\bm x}_1,t) \eqspace ,
\end{equation}
which describes the fact that if the separation between two points becomes infinitesimally small, the vorticities $\omega_1$ and $\omega_2$ should also converge to the same value, i.\,e.~the vorticity field is smooth on small length scales.

Beside the joint PDFs we can define the {\em conditional} PDF
\begin{equation}
  p(\omega_1,{\bm x}_1|\omega_2,{\bm x}_2, \dots , \omega_N,{\bm x}_N,t) = \frac{f_N(\omega_1,{\bm x}_1,\omega_2,{\bm x}_2, \dots , \omega_N,{\bm x}_N,t)}{f_{N-1}(\omega_2,{\bm x}_2, \dots , \omega_N,{\bm x}_N,t)} 
\end{equation}
which is the probability of finding $\omega_1$ at ${\bm x}_1$ for  given $\omega_2,\dots,\omega_{N}$ at ${\bm x}_2,\dots,{\bm x}_{N}$.
Now we can also define the conditional averages
\begin{equation}
  \langle \omega({\bm x}_1,t)| \omega_2,{\bm x}_2, \dots , \omega_N,{\bm x}_N,t \rangle = \int \! \mathrm{d} \omega_1 \, \omega_1 \, p(\omega_1,{\bm x}_1|\omega_2,{\bm x}_2, \dots , \omega_N,{\bm x}_N,t) 
\end{equation}
which allows us to write down relations such as (see Appendix H in \cite{pope00book} for details)
\begin{align}\label{eq:condAvFineGrained}
   \langle \omega({\bm x}_1,t)&\hat{f}_{N-1}( \omega_2,{\bm x}_2,\dots , \omega_N,{\bm x}_N,t) \rangle \nonumber \\ &= \langle \omega({\bm x}_1,t)| \omega_2,{\bm x}_2, \dots , \omega_N,{\bm x}_N,t \rangle f_{N-1}(\omega_2,{\bm x}_2, \dots , \omega_N,{\bm x}_N,t)
\end{align}
that will play a crucial role for the derivation of the LMN hierarchy.

To keep the notation as compact as possible we will from now on skip the indices of the PDFs, because the order of the PDF becomes clear by its arguments.
Additionally we introduce the abbreviation  
\begin{equation}
  f(\lbrace \omega_l,{\bm x}_l \rbrace,t) := f_N(\omega_1,{\bm x}_1, \dots , \omega_N,{\bm x}_N,t)
\end{equation}
for the $N$-point PDF where $\lbrace \omega_l,{\bm x}_l \rbrace$ denotes the set of $N$ points $\omega_1,{\bm x}_1, \dots , \omega_N,{\bm x}_N$.
In the following we will also  frequently use the notation $f(\omega',{\bm x}',\lbrace \omega_l,{\bm x}_l \rbrace,t)$ for the $N+1$-point PDF.

Since conditional averages like e.\,g.~in eq.~\eqref{eq:condAvFineGrained} are central quantities, it would be helpful to derive some general analytical expressions for them.
In case the characteristic function of the $N+1$-point distribution 
$f(\omega',{\bm x}',\lbrace \omega_l,{\bm x}_l \rbrace,t)$ is known the conditional averages can be expressed by 
\begin{align}\label{eq:analyticCond}
  \langle  \omega',\bm{x}'|\lbrace \omega_l,{\bm{x}}_l\rbrace \rangle f(\lbrace \omega_l,{\bm x}_l \rbrace,t) &= \nonumber \\
  -\bigg[\frac{1}{\mathrm{i}} \frac{\partial}{\partial \alpha'} W\bigg(&\alpha',\bm{x}',\bigg\lbrace -\frac{1}{\mathrm{i}} \frac{\partial }{\partial \omega_l} ,{\bm{x}}_l  \bigg\rbrace\bigg)\bigg]_{\alpha'=0} f(\lbrace \omega_l,{\bm x}_l \rbrace,t)
\end{align}
where  $W(\alpha',\bm{x}',\lbrace \alpha_l ,{\bm{x}}_l \rbrace)$ is the negative logarithm of the characteristic function known as \textit{cumulant generating function}.
For a multivariate Gaussian PDF the conditional expectation can be expressed in terms of the correlation functions.
The result is 
\begin{equation}\label{analyticCondGauss}
  \langle  \omega',\bm{x}'|\lbrace \omega_l,{\bm{x}}_l\rbrace \rangle  =  \sum\limits_{k,m} C(\bm{x}'-{\bm{x}}_k) C^{-1}({\bm{x}}_k-{\bm{x}}_m) \omega_m
\end{equation}
where $C(\bm{x}'-{\bm{x}}_k)$ stands for the two-point vorticity correlation functions, and $C^{-1}(\bm{x}_i-\bm{x}_j)$ denotes the $(i,j)$-th element of the inverse of the $N\times N$ correlation matrix $\mathrm{C}$ that characterizes the multivariate Gaussian.
We refer the interested reader to ref.~\cite{Friedrich2010arx} for a derivation of \eqref{eq:analyticCond} and \eqref{analyticCondGauss}.

\subsection{Statistical Symmetries}
From the observation of turbulent flows it is immediately clear that individual realizations of turbulent fields break spatio-temporal symmetries.
Depending on the flow under consideration, however, symmetries may be recovered in a statistical sense \cite{frisch95book}, i.\,e.~due to the averaging procedure.
The rigorous study of statistical symmetries has been introduced to hydrodynamical turbulence by Robertson \cite{Robertson1940}, Chandrasekhar \cite{Chandrasekhar1951,Chandrasekhar1951B} and Batchelor \cite{batchelor53book} and since then has been an indispensable mathematical tool for the statistical investigation of turbulent flows.
The Howarth-K\'{a}rm\'{a}n relation \cite{Karman1938prs} may serve as a prominent example here demonstrating that statistical symmetries may not only help to simplify the statistical equations, but also may cast them into a tractable mathematical form allowing for new insights.
Of course this is also the case for the application of statistical symmetries to kinetic equations, and in fact, some of the results presented in the following depend crucially on the exploitation of these symmetries.

Among the simplest of these symmetries are statistical stationarity, homogeneity and isotropy, and each of these symmetries can be defined as the invariance of a given statistical quantity with respect to a symmetry transformation.
In case of stationarity and homogeneity, the corresponding transformations are temporal and spatial shifts, respectively.
For the $N$-point vorticity PDF of two-dimensional turbulence this leads to the following relations 
\begin{subequations}
\begin{align}
  f(\{\omega_{l},\bm{x}_{l}\},t) &=  f(\{\omega_{l},\bm{x}_{l}+\bm{r}\},t)\\
  f(\{\omega_{l},\bm{x}_{l}\},t) &=  f(\{\omega_{l},\bm{x}_{l}\},t+\tau)\eqspace,
\end{align}
\end{subequations}
where $\bm{r}$ and $\tau$ are arbitrary temporal and spatial shifts.
Isotropy is defined as invariance with respect to rotations and (optionally) reflections.
For the $N$-point velocity PDF this leads to 
\begin{equation}
  f\left(\{\bm{u}_{l},\bm{x}_{l}\},t\right)=f\left(\left\{ \mathrm{R}\bm{u}_{l},\mathrm{R}\bm{x}_{l}\right\} ,t\right) \eqspace ,\label{eq:isotropy-trafo-pdf}
\end{equation}
where $\mathrm{R}$ is an arbitrary rotation matrix.
For the vorticity PDF one has to take into account that the vorticity is a pseudo-scalar, i.\,e.~it changes sign when $\mathrm{R}$ contains a reflection.

To demonstrate the importance of these symmetries let us consider the single-point velocity PDF $f(\bm{u},\bm{x},t)$ in homogeneous, isotropic and stationary three-dimensional turbulence.
As the PDF depends only on one spatial and one temporal point, the invariance under the corresponding shifts implies independence of these variables, i.\,e.~$f(\bm{u},\bm{x},t)=f(\bm{u})$.
The invariance under rotations, i.\,e.~$f(\mathrm{R}\bm{u})=f(\bm{u})$, implies that the PDF can only depend on the magnitude of the velocity vector instead of the three vector components.
Counting all of these simplifications together, we have reduced the dependency of the single-point PDF from seven (one temporal, three spatial and three sample-space) variables to one, the velocity magnitude.

The fact that a PDF does not depend on certain variables (due to statistical symmetries) allows one to formulate  a whole sequence of relations which involve conditional expectations.
Let us exemplify this for the PDF of a field $\omega({\bm x},t)$ with stationary statistics.
The time derivative as well as the higher-order time derivatives have to vanish identically.
This leads to
\begin{align}
  0 = \frac{\partial }{\partial t}f(\omega,{\bm x})= &
  -\frac{\partial }{\partial \omega}\biggl\langle \left[\frac{\partial}{\partial t} \omega({\bm x},t)\right]
  \delta\bigl(\omega-\omega({\bm x},t)\bigr) \biggr\rangle \nonumber\\
  =& -\frac{\partial }{\partial \omega}\biggl\langle\frac{\partial}{\partial t} \omega({\bm x},t) \bigg| \omega \biggr\rangle
  f(\omega,{\bm x})\label{eq:fdot}\eqspace.
\end{align}
Here we already encounter the closure problem, from eq.~\eqref{eq:fdot} it is clear that we need the joined PDF of $\frac{\partial}{\partial t}\omega$ and $\omega$ to determine the temporal variation of $f(\omega,{\bm x},t)$.
The resulting equation for $\frac{\partial^2}{\partial t^2} f(\omega,{\bm x})=0$ has been
highlighted by Pope and Ching \cite{pope1993pof}:
\begin{equation}
  0=-\frac{\partial }{\partial \omega}
  \biggl\langle \frac{\partial^2}{\partial t^2} \omega({\bm x },t)\bigg|\omega \biggr\rangle f(\omega,{\bm x})
  +\frac{\partial^2 }{\partial \omega^2}
  \biggl\langle \left[\frac{\partial}{\partial t}\omega({\bm x },t)\right]^2\bigg|\omega \biggr\rangle f(\omega,{\bm x})
\end{equation}
Obviously this treatment can be generalized to all continuous symmetries.
For translationally invariant statistics the simplest relation reads
\begin{equation}
  0=-\frac{\partial }{\partial \omega}
  \langle \Delta \omega({\bm x },t)|\omega \rangle f(\omega,t)
  +\frac{\partial^2 }{\partial \omega^2}
  \langle [\nabla \omega({\bm x },t)]^2|\omega \rangle 
  f(\omega,t)  \eqspace .
\end{equation}
This type of relation will play an important role in the treatment of dissipative terms in the evolution equations, see section~\ref{sec:singlepointvelocity}.

Up to now we have applied the symmetries to PDFs, which are scalar-valued functions.
However, the kinetic equations of the LMN hierarchy also contain vector- and tensor-valued functions, e.\,g.~$\left\langle \left.\Delta\bm{u}\right|\bm{u}\right\rangle $.
Their representation can also be simplified with the help of statistical symmetries, which leads to the general concept known as the invariant theory of tensors.
Generally speaking, exploiting statistical symmetries, vectors and tensors can be expressed as a superposition of a generating set of covariant vectors or tensors.

Here we would like to exemplify this concept for isotropic vector- and tensor-valued functions depending on one vector variable.
Such functions appear in the context of the LMN hierarchy as conditional averages of vectors and tensors conditioned on a vector, e.\,g.~$\left\langle \left.\Delta\bm{u}\right| \bm{u}\right\rangle $ and $\langle\nabla\bm{u}\left(\nabla\bm{u}\right)^{T}|\bm{u}\rangle $.
Due to the fact that in two dimensions fewer reductions are possible we restrict the further discussion to three dimensions.

Let $\left\langle \left.\bm{a}\right|\bm{u}\right\rangle $ and $\left\langle \left.B_{ij}\right|\bm{u}\right\rangle $ be a conditionally averaged vector and symmetric tensor in three dimensions.
Statistical isotropy demands that they obey the transformation rules
\begin{subequations}
\begin{align}
  \left\langle \left.\bm{a}\right|{\rm R}\bm{u}\right\rangle  & =  \left\langle \left.{\rm R}\bm{a}\right|\bm{u}\right\rangle \\
  \left\langle \left.{\rm B}\right|{\rm R}\bm{u}\right\rangle  & =  \left\langle \left.{\rm RBR^{T}}\right|\bm{u}\right\rangle \eqspace,
\end{align}
\end{subequations}
which can be derived from the transformation rule of a PDF \eqref{eq:isotropy-trafo-pdf}.
Due to these symmetries this conditional averages must have the representations
\begin{subequations}
\begin{align}
  \left\langle \left.\bm{a}\right|\bm{u}\right\rangle  & =  \left\langle \left.\hat{\bm{u}}\cdot\bm{a}\right|u\right\rangle \hat{\bm{u}}\\
  \left\langle \left.B_{ij}\right|\bm{u}\right\rangle  & =  \mu(u)\delta_{ij}+\left[\lambda(u)-\mu(u)\right]\hat{u}_{i}\hat{u}_{j}\\
  \lambda(u) & =  \left\langle \left.\hat{\bm{u}} \cdot {\rm B}\hat{\bm{u}}\right|u\right\rangle \\
  \mu(u) & =  \frac{1}{2}\left\langle \left.{\rm Tr}\left({\rm B}\right)\right|u\right\rangle -\frac{1}{2}\left\langle \left.\hat{\bm{u}}\cdot {\rm B}\hat{\bm{u}}\right|u\right\rangle \eqspace,
\end{align}
\end{subequations}
where $\hat{\bm{u}}=\bm{u}/u$ is the unit vector in direction of $\bm{u}$.
These results can be rigorously derived, where the core idea is to use rotations around the axis $\hat{\bm{u}}$.

These representations have several advantages.
On the one hand they make it possible to simplify the equations containing this conditional averages.
On the other hand the symmetry reduction renders the estimation of the conditional averages from DNS feasible as only scalars conditioned on scalars have to be estimated from the data, in contrast to vectors or tensors conditioned on vectors.

Finally, we would like to mention here that statistical symmetries can be used to formulate further constraints on the conditional averages like $\left\langle \left.\hat{\bm{u}}\cdot\bm{a}\right|u\right\rangle $ and $\lambda(u)$.
For example, $\lambda(u)$ has to be an even function in $u$, i.\,e.~its Taylor series may contain only even powers of $u$.
For further details we refer to \cite{wilczek11jfm}.

\subsection{LMN Hierarchy for the Vorticity in Two-Dimensional Turbulence}\label{lmn_2d}
We now proceed to a detailed presentation of the hierarchy of evolution equations for the PDFs.
To keep things simple, we focus on the hierarchy for the vorticity field in two-dimensional turbulence.
The corresponding, analogous hierarchies for the three-dimensional case will be summarized later.

We start from the evolution equation for vorticity $\boldsymbol{\omega}=\omega {\bm e}_z$, where the vorticity component $\omega({\bm x},t)$ depends on the horizontal coordinates ${\bm x}=(x_1,x_2)$.
The basic fluid dynamic equation for the vorticity reads
\begin{equation}\label{eq:2dvorticity}
  \frac{\partial}{\partial t} \omega({\bm x},t)+{\bm u}({\bm x},t)\cdot \nabla_{\bm x}\, \omega({\bm x},t)=L(-\Delta_{\bm x})\omega({\bm x},t)+F({\bm x},t) \eqspace .
\end{equation}
We have included a forcing term, $F({\bm x},t)$ as well as a general viscous term, $L(-\Delta)\omega({\bm x},t)$, which may contain large-scale friction, usual dissipation and hyperviscous terms.
A suitable choice of $F$ and $L$ allows one to generate a direct, an inverse \cite{Friedrich2010arx} as well as dual cascades (see e.\,g.~\cite{boffetta2010pre}).
The velocity field can be determined using Biot-Savart's law, which relates the vorticity field to the velocity field:
\begin{equation}\label{eq:biot_savart}
  {\bm u}({\bm x},t) =\int \! \di {\bm x}' \, \omega({\bm x}',t) {\bm e}_z \times 
  \frac{{\bm x}-{\bm x}'}{2\pi|{\bm x}-{\bm x}'|^2} =\int \! \di {\bm x}' \, {\bm U}({\bm x}-{\bm x}')\omega({\bm x}',t)
\end{equation}
The velocity field ${\bm U}({\bm x}-{\bm x}')$ may be regarded as the field of a single point vortex located at ${\bm x}'$.
It is purely azimuthal and decays like $1/|{\bm x}-{\bm x}'|$.

We now want to proceed to a statistical description of the vorticity.
To derive the kinetic equation for the single-point statistics of two-dimensional turbulence, we can use the concepts introduced in section~\ref{sec:concepts} in a straightforward manner.
To this end we consider the fine-grained PDF
\begin{equation}
  \hat f_1(\omega_1,{\bm x}_1,t) =\delta\bigl(\omega_1-\omega({\bm x}_1,t)\bigr)
\end{equation}
and take the time derivative, which leads to
\begin{equation}
  \frac{\partial}{\partial t} \hat f_1(\omega_1,{\bm x}_1,t) = \left[ -\frac{\partial}{\partial \omega_1} \hat f_1(\omega_1,{\bm x}_1,t)\right] \frac{\partial}{\partial t}\omega({\bm x}_1,t) \eqspace .
\end{equation}
Analogously we compute
\begin{equation}
  \bm u(\bm x_1,t) \cdot \nabla_{\bm x_1} \hat f_1(\omega_1,{\bm x}_1,t) = \left[ -\frac{\partial}{\partial \omega_1} \hat f_1(\omega_1,{\bm x}_1,t)\right] \bm u(\bm x_1,t) \cdot \nabla_{\bm x_1} \omega({\bm x}_1,t) \eqspace .
\end{equation}
Adding these results we obtain
\begin{align}
  &\frac{\partial}{\partial t} \hat f_1(\omega_1,{\bm x}_1,t) + \bm u(\bm x_1,t) \cdot \nabla_{\bm x_1} \hat f_1(\omega_1,{\bm x}_1,t) \label{eq:1pderivationa} \\
  &= \frac{\partial}{\partial t} \hat f_1(\omega_1,{\bm x}_1,t) + \nabla_{\bm x_1}\cdot [\bm u(\bm x_1,t) \hat f_1(\omega_1,{\bm x}_1,t)] \label{eq:1pderivationb} \\
  &= \left[ -\frac{\partial}{\partial \omega_1} \hat f_1(\omega_1,{\bm x}_1,t)\right] \left[ \frac{\partial}{\partial t}\omega({\bm x}_1,t) + \bm u(\bm x_1,t) \cdot \nabla_{\bm x_1} \omega({\bm x}_1,t) \right] \label{eq:1pderivationc} \\
  &= \left[ -\frac{\partial}{\partial \omega_1} \hat f_1(\omega_1,{\bm x}_1,t)\right] \left[ L(-\Delta_{\bm x_1})\omega({\bm x_1},t)+F({\bm x_1},t) \right] \label{eq:1pderivationd} \\
  &= -\frac{\partial}{\partial \omega_1} \left[ L(-\Delta_{\bm x_1})\omega({\bm x_1},t)+F({\bm x_1},t) \right] \hat f_1(\omega_1,{\bm x}_1,t)\eqspace. \label{eq:1pderivatione}
\end{align}
For eq.~\eqref{eq:1pderivationb} we have used solenoidality of the velocity field.
From eq.~\eqref{eq:1pderivationc} to eq.~\eqref{eq:1pderivationd} we have replaced the left-hand side of the vorticity equation \eqref{eq:2dvorticity} with its right-hand side.
The step from eq.~\eqref{eq:1pderivationd} to eq.~\eqref{eq:1pderivatione} finally makes use of the fact that the terms of the vorticity equation are independent of the sample-space variable, which concludes the derivation of the single-point equation for the fine-grained PDF.
Ensemble averaging then leads to the desired PDF equation,
\begin{align}
  \frac{\partial}{\partial t} f_1(\omega_1,{\bm x_1},t) &+ \nabla_{\bm x_1} \cdot \langle \bm u(\bm x_1,t) \hat f_1(\omega_1,{\bm x}_1,t) \rangle = \\
  &-\frac{\partial}{\partial \omega_1} \langle \left[ L(-\Delta_{\bm x_1})\omega({\bm x}_1,t)+F({\bm x_1},t) \right] \hat f_1(\omega_1,{\bm x}_1,t)\rangle \eqspace ,
\end{align}
which confronts us with the closure problem in terms of the joint averages of the fine-grained PDF and velocity, the dissipative term and the forcing term.
To proceed, two options are available.
First, the joint averages can be expressed as conditional averages according to
\begin{align}
  \langle \bm u(\bm x_1,t) \hat f_1(\omega_1,{\bm x}_1,t) \rangle &= \langle \bm u(\bm x_1,t) | \omega_1, \bm x_1, t  \rangle f_1(\omega_1,{\bm x}_1,t) \\
  \langle [L(-\Delta_{\bm x_1})\omega({\bm x}_1,t)] \hat f_1(\omega_1,{\bm x}_1,t) \rangle &= \langle L(-\Delta_{\bm x_1})\omega({\bm x}_1,t) | \omega_1, \bm x_1, t \rangle f_1(\omega_1,{\bm x}_1,t) \\
  \langle F({\bm x_1},t) \hat f_1(\omega_1,{\bm x}_1,t) \rangle &= \langle F({\bm x_1},t) | \omega_1, \bm x_1, t \rangle f_1(\omega_1,{\bm x}_1,t) \eqspace .
\end{align}
These terms are in principle accessible by DNS data, which will be discussed in detail below.
The second option is to express the unclosed velocity and dissipative terms by the two-point statistics.
For the velocity term we obtain by insertion of the identity $\int \! \di \omega_2 \, \delta\bigl(\omega_2 - \omega(\bm x_2,t)\bigr)=1$
\begin{equation}
  \langle {\bm u}({\bm x}_1,t) \hat f(\omega_1,{\bm x}_1,t)\rangle =\int \! \di {\bm x}_2 \int \! \di \omega_2 \, {\bm U}({\bm x}_1-{\bm x}_2) \, \omega_2 \,  f(\omega_1,{\bm x}_1, \omega_2,{\bm x}_2,t) \eqspace .
\end{equation} 
At this stage the coupling to the two-point PDF arises, in much the same way as for the Friedmann-Keller hierarchy for the correlation functions.
A similar treatment is then performed for the dissipative term, which leads to
\begin{align}
  \langle [L(-\Delta_{{\bm x}_1}) & \omega({\bm x}_1,t)] \hat f(\omega_1,{\bm x}_1,t)\rangle \nonumber \\
  &=\int \! \di {\bm x}_2 \int \! \di \omega_2 \, \delta\bigl({\bm x}_2-{\bm x}_1\bigr) \, \omega_2 \, L(-\Delta_{{\bm x}_2})  f(\omega_1,{\bm x}_1, \omega_2,{\bm x}_2,t) \eqspace . \nonumber
\end{align}
Up to now, we have not further specified the forcing term.
For a deterministic forcing term, the unclosed term simply involves the joint statistics of the external force field and the vorticity field.
Especially for two-dimensional turbulence a white-in-time forcing term with Gaussian statistics is often considered, which can be treated further.
Such a forcing term is fully specified by the correlation function
\begin{equation}
  \langle F({\bm x}_1,t)F({\bm x}_2,t')\rangle =Q({\bm x}_1-{\bm x}_2) \delta\bigl(t-t'\bigr) \eqspace .
\end{equation}
 The transition from the fine-grained PDF to the full PDF then involves an averaging procedure with respect to the white noise force field $F({\bm x},t)$.
 This yields a representation of the forcing terms as
\begin{equation}
  \langle F({\bm x}_1,t) \hat  f(\bm \omega_1,\bm x_1, \bm \omega_2,\bm x_2,t) \rangle =-\frac{1}{2} \sum_{j=1}^2 Q({\bm x}_1-{\bm x}_j) \frac{\partial }{\partial \omega_j} f(\bm \omega_1,\bm x_1, \bm \omega_2,\bm x_2,t) \eqspace .
\end{equation}
The derivation of this result is similar to the derivation of the Fokker-Planck equation for a set of Langevin equations \cite{Risken96book,Gardiner04book}.
The random force method applied to turbulence has also been discussed by Novikov \cite{novikov65spj}.

The whole derivation is generalized to the $N$-point statistics in a straight-forward manner by starting from the $N$-point fine-grained PDF.
Expressing the unclosed terms in terms of the $(N+1)$-statistics then results in
\begin{align}
  \frac{\partial }{\partial t}f(\lbrace \omega_l,{\bm x}_l \rbrace,t) 
  &+\sum_{j=1}^N \nabla_{{\bm x}_j} \cdot \int \! \di \omega' \di {\bm x}' \, {\bm U}({\bm x}_j-{\bm x}')\omega' f(\omega',{\bm x}',\lbrace \omega_l,{\bm x}_l \rbrace,t) \nonumber \\
  = & -\sum_{j=1}^N \frac{\partial }{\partial \omega_j} \int \! \di \omega' \di {\bm x}' \, \delta\bigl({\bm x}_j-{\bm x}'\bigr) L(-\Delta_{{\bm x}'}) \omega' f(\omega',{\bm x}',\lbrace \omega_l,{\bm x}_l \rbrace,t) \nonumber \\
  & -\sum_{j=1}^N \frac{\partial }{\partial \omega_j} \langle F({\bm x}_j,t) \hat f(\omega',{\bm x}',\lbrace \omega_l,{\bm x}_l \rbrace,t) \rangle \eqspace ,
\end{align}
where again the coupling to the $(N+1)$-point PDFs arises due to the nonlocal Biot-Savart law relating velocity and vorticity field, as well as the dissipative term.
The alternative formulation is obtained by replacing the terms containing the $(N+1)$-point probability functions by the corresponding conditional averages.
The set of kinetic equations then takes the form
\begin{align}\label{eq:Npointwithconave}
  \frac{\partial }{\partial t} f(\lbrace \omega_l,{\bm x}_l \rbrace,t) &+
  \sum_{j=1}^N \nabla_{{\bm x}_j} \cdot \langle {\bm u}({\bm x}_j,t)|\lbrace \omega_l,{\bm x}_l \rbrace\rangle f(\lbrace \omega_l,{\bm x}_l \rbrace,t) \nonumber \\
&= -\sum_{j=1}^N \frac{\partial }{\partial \omega_j}
\langle \mu(\bm x_j,t) |\lbrace \omega_l,{\bm x}_l \rbrace \rangle
f(\lbrace \omega_l,{\bm x}_l \rbrace,t) \eqspace ,
\end{align}
where we have introduced the abbreviation for the right-hand side of the vorticity equation \eqref{eq:2dvorticity}
\begin{equation}
  \mu(\bm x_j,t) = L(-\Delta_{{\bm x_j}})\omega({\bm x_j},t) + F({\bm x_j},t) \eqspace .
\end{equation}
In this formulation, the chain of PDF equations is retained up to the order $N$ and formally truncated on this level by the introduction of the up to this point unspecified conditional averages.
The case of $N=2$ will be discussed with the input from DNS data in section~\ref{sec:2dlmndns}.

\subsection{The Method of Characteristics} \label{sec:method_of_characteristics}
The kinetic equations formulated in terms of conditional expectations formally are first-order partial differential equations.
Hence, they can be solved using the method of characteristics~\cite{courant62book}.
By application of this method, one can identify trajectories in the phase space spanned by the variables of the partial differential equation, the so-called characteristic curves or simply \emph{characteristics}.
Along these characteristics, the original partial differential equation transforms into an ordinary differential equation.

Let us exemplify the procedure for the simplest case of the kinetic equations for two-dimensional flows, i.\,e.~the $N$-point kinetic equation \eqref{eq:Npointwithconave}.
The associated characteristics are solutions of the equations 
\begin{subequations}
\label{eq:moc_ode_char}
\begin{align}
  \frac{\di}{\di s}t(s) &= 1  \label{eq:moc_ode_char_t} \\
  \frac{\di}{\di s}{\bm X}_j(s) &= \langle {\bm u}({\bm x}_j,t)|\lbrace \omega_l,{\bm x}_l \rbrace \rangle_{\{\omega_l,{\bm x}_l\}=\{\Omega_l(s),{\bm X}_l(s)\}} \\ 
  \frac{\di}{\di s}\Omega_j(s) &= \langle \mu({\bm x}_j,t)|\lbrace \omega_l, {\bm x}_l \rbrace \rangle_{\{\omega_l,{\bm x}_l\}=\{\Omega_l(s),{\bm X}_l(s)\}} \eqspace ,
\end{align}
\end{subequations}
where $s$ is the parametrization of the trajectories through phase space, and the conditional averages have to be evaluated at the current positions along the characteristics.
The solutions of these ordinary differential equations are the characteristics ${\bm X}_j(s,\lbrace \omegalzero,\xlzero \rbrace)$ and $\Omega_j(s,\lbrace \omegalzero,\xlzero \rbrace)$ that are depending on the initial conditions $\{\omegalzero,\xlzero\}$ at $s=0$:
\begin{subequations}
\begin{align}
  {\bm X}_j(0,\lbrace \omegalzero,\xlzero \rbrace) =\xjzero\\
  \Omega_j(0,\lbrace \omegalzero,\xlzero \rbrace)=\omegajzero
\end{align}
\end{subequations}
Along these characteristics, the kinetic equation \eqref{eq:Npointwithconave} transforms into an ordinary differential equation:
\begin{align}
  \frac{\di}{\di s} f(s) = -\sum_{j=1}^N \Bigl[&\nabla_{\bm x_j}\cdot  \langle {\bm u}({\bm x}_j,t)|\lbrace \omega_l,{\bm x}_l \rbrace \rangle \nonumber \\
  &+ \frac{\partial }{\partial \omega_j} \langle \mu({\bm x}_j,t)|\lbrace \omega_l, {\bm x}_l \rbrace\rangle\Bigr]_{\{\omega_l,{\bm x}_l\}=\{\Omega_l(s),{\bm X}_l(s)\}}\, f(s) 
\end{align}
This means that the deformation of the PDF along the characteristics is governed by the bracketed term on the right-hand side, which is the phase space divergence of the vector field formed by the conditional averages appearing in eq.~\eqref{eq:moc_ode_char}.

If the PDF at the initial point $\lbrace \omegalzero,\xlzero\rbrace$ of the characteristic curve, $f_0(\lbrace \omegalzero,\xlzero \rbrace)$, is known, the PDF at ``later'' points ${\bm x}_j={\bm X}_j(s,\lbrace \omegalzero,\xlzero \rbrace)$, $\omega_j=\Omega_j(s,\lbrace \omegalzero,\xlzero \rbrace)$ (i.\,e.~for a certain value of $s>0$) is determined by 
\begin{equation}
  f(\lbrace\omega_l,{\bm x}_l\rbrace)= J\,f_0(\lbrace \omegalzero,\xlzero \rbrace) \eqspace .
\end{equation}
Here, $J$ denotes the Jacobian determinant of the inverse mapping $\bigl({\bm X}_j^{-1},\Omega_j^{-1}\bigr)$ that matches the phase space points ${\bm x}_j$ and $\omega_j$ back to the initial conditions of the characteristic that passes through them, i.\,e.
\begin{subequations}
\begin{align}
  {\bm X}_j^{-1}(\{{\bm x}_l,\omega_l \}) &= \xjzero \\
  \Omega_j^{-1}(\{{\bm x}_l,\omega_l \}) &= \omegajzero \eqspace .
\end{align}
\end{subequations}

The method of characteristics exhibits some interesting analogies to the Lagrangian description of turbulence.
To this end one can note that evolution equations of the characteristic curves, eqs.~\eqref{eq:moc_ode_char}, describe the motion of Lagrangian fluid particles that are moving within the conditionally averaged velocity field and change their vorticity according to the conditionally averaged dissipation and forcing fields.
In this sense the characteristic equations describe the average dynamics of a class of fluid particles defined by the configuration of the $N$ fluid particles under consideration.
This view is closely related to the concept of quasi-particles.

Furthermore, in the continuum limit $N\rightarrow\infty$ or $\bigl({\bm X}_j,\Omega_j\bigr) \rightarrow \bigl({\bm X}({\bm y},t), \Omega({\bm y},t)\bigr)$, where the characteristics are identified by their continuous starting positions ${\bm y}$ that cover the whole space, the characteristic equations tend to the vorticity equation formulated in the Lagrangian framework,
\begin{subequations}
\begin{align}
  \frac{\mathrm{d} }{\mathrm{d} t} {\bm X}({\bm y},t) &= \int \! \di{\bm y}'\, \Omega({\bm y}',t)\, {\bm U}\bigl({\bm X}({\bm y},t)-{\bm X}({\bm y}',t)\bigl) \\
  \frac{\mathrm{d} }{\mathrm{d} t}\Omega({\bm y},t) &= \bigl[L(-\Delta) \omega({\bm x},t)+F({\bm x},t)\bigr]_{{\bm x}={\bm X}({\bm y},t)} \eqspace ,
\end{align}
\end{subequations}
with ${\bm U}$ again being the Biot-Savart kernel.
The formulation, modeling, and investigation of the characteristic equations, therefore, allows one to make contact with models for the motion of few Lagrangian tracers, especially tetrad models \cite{chertkov99pof,naso05pre} and models for the velocity gradient tensor \cite{chevillard06prl,chevillard08pof}.
We refer to the reviews of Pumir and Naso, and Li in this volume.

\subsection{From the Eulerian to the Lagrangian Frame}
The discussion of the preceding section links our discussion of the LMN hierarchy to the Lagrangian description of turbulence.
An extended discussion of the Lagrangian approach would be beyond the scope of this review paper but we want to make at least some comments on this issue.
The choice of the frame of reference depends mainly on the physical problem under investigation.
Whereas the Eulerian picture seems for example preferable for questions connected to the energy flux in scale, the Lagrangian picture is the natural way to describe transport and mixing processes.
To this end one considers the Lagrangian path, ${\bm X}({\bm y},t)$, of a Lagrangian particle starting at initial point ${\bm y}$ at time $t=0$.
Furthermore, one introduces the Lagrangian velocity field or vorticity field according to ${\bm u}_L({\bm y},t)={\bm u}({\bm X }({\bm y},t),t)$, $\boldsymbol{\omega}_L({\bm y},t)=\boldsymbol{\omega}({\bm X }({\bm y},t),t)$.
The Lagrangian single-point PDF is then defined as
\begin{equation}\label{Lagpdf}
  f_L({\bm x},{\bm u},{\bm y},t)=\left\langle \delta\bigl({\bm x}-{\bm X}({\bm y},t)\bigr) \delta\bigl({\bm u}-{\bm u}_L({\bm y},t)\bigr) \right\rangle
\end{equation}
with a straightforward extension to $N$ particles.
As for the Eulerian case one can now proceed to construct a LMN hierarchy for the Lagrangian quantities of interest.
For example the chain of equations for the case of the velocity PDF has been formulated in \cite{Friedrich2003prl,Friedrich2002arx}, whereas the case of the vorticity PDF has been investigated in \cite{Novikov1997pla}.

Another important topic is connected to the general relation between the Lagrangian and the Eulerian picture.
In some situations it is helpful to understand the statistics of an observable in one frame of reference given the statistics of this observable in the other frame of reference.
One famous example is the translation of the two-point two-time velocity autocorrelation function from the Eulerian into the Lagrangian picture (see e.\,g.~\cite{Ott05njp} and references therein).
In recent years especially the statistics of Eulerian and Lagrangian velocity increments have gained much interest.
Experiments and numerical simulations showed that the deviation from Gaussianity of these PDFs is in the Lagrangian frame even stronger than in the Eulerian frame.
Attempts to relate increment PDFs and structure functions have been made in Refs.~\cite{Borgas1993pt, Boffetta2002pre, Chevillard2003prl} mainly based on dimensional arguments.

In \cite{Kamps2009pre} the formalism of fine-grained distributions introduced in section~\ref{sec:concepts} was used to derive an exact relation of these quantities which includes the transition relations used in \cite{Borgas1993pt, Boffetta2002pre, Chevillard2003prl} as a special case.
The main result of \cite{Kamps2009pre} is that in general one has to add information in form of transition PDFs to translate the PDFs of an observable from one frame of reference to the other.
The resulting relation that translates the increment PDFs involves two transition PDFs which add the missing information.
One takes into account the dispersion of the Lagrangian tracers and the other adds information on the time development of the flow field at the starting point of the tracer.
With the help of this relation it was for example possible to understand how the nearly Gaussian PDFs of the Eulerian velocity increments in the inverse cascade of two-dimensional turbulence translate into  non-Gaussian PDFs in the Lagrangian frame.
An application to three-dimensional hydro- and magneto-hydrodynamic turbulence can be found in \cite{Homann2009njp}.
In principle the methods to derive the results of \cite{Kamps2009pre} can be used in general to find exact relations between Eulerian and Lagrangian PDFs for various observables.

\section{Inverse Cascade in Two-dimensional Turbulence} \label{sec:2dlmndns}
\begin{figure}[t]
 \centering{\includegraphics[width=0.7\textwidth]{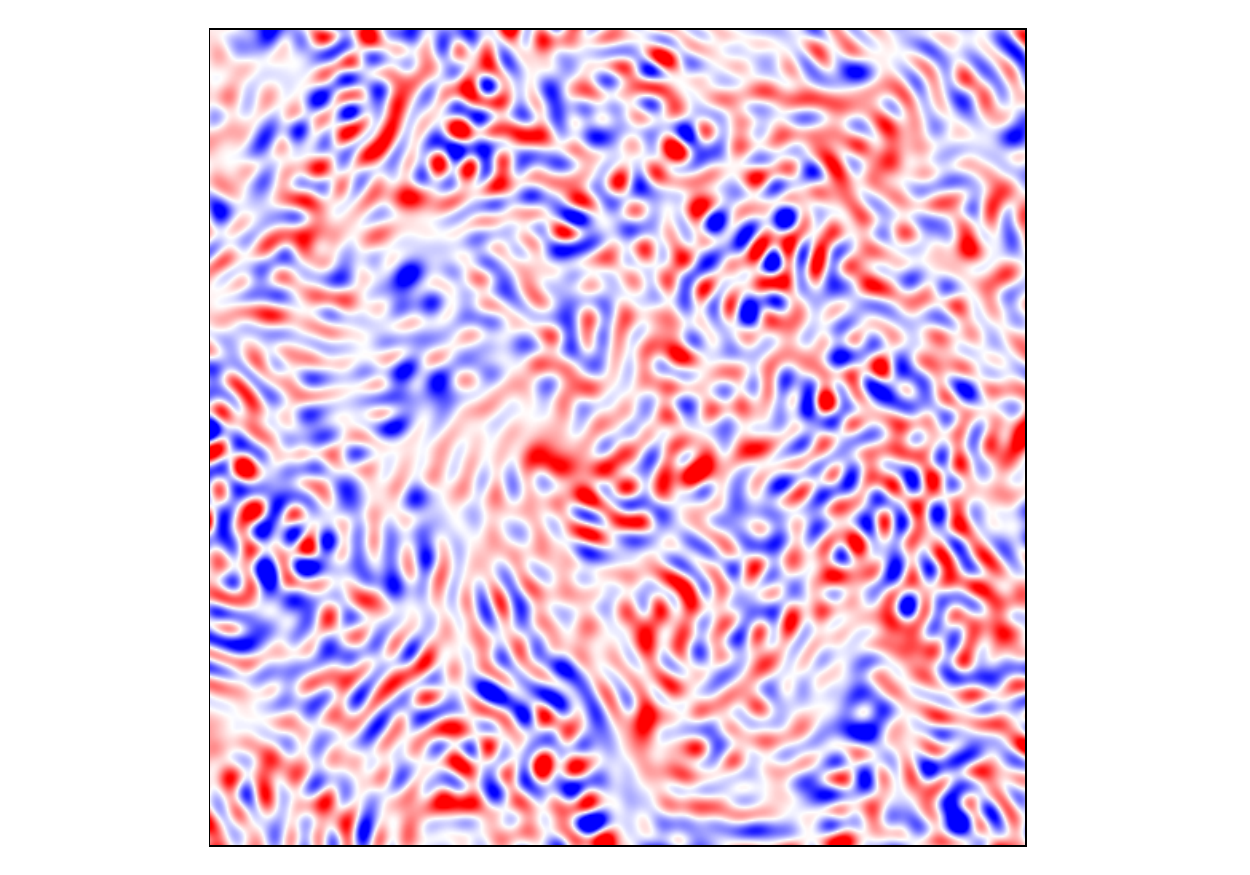}}
 \caption{Visualization of the vorticity field in two-dimensional turbulence in the inverse cascade regime. The vorticity field is taken from simulation A in \cite{Friedrich2010arx}. Videos of this visualization can be found in \cite{url:turbulenceteamms}.}
 \label{fig:2d_vorticity_viz}
\end{figure}

\begin{figure}[t]
 \includegraphics[width=0.5\textwidth]{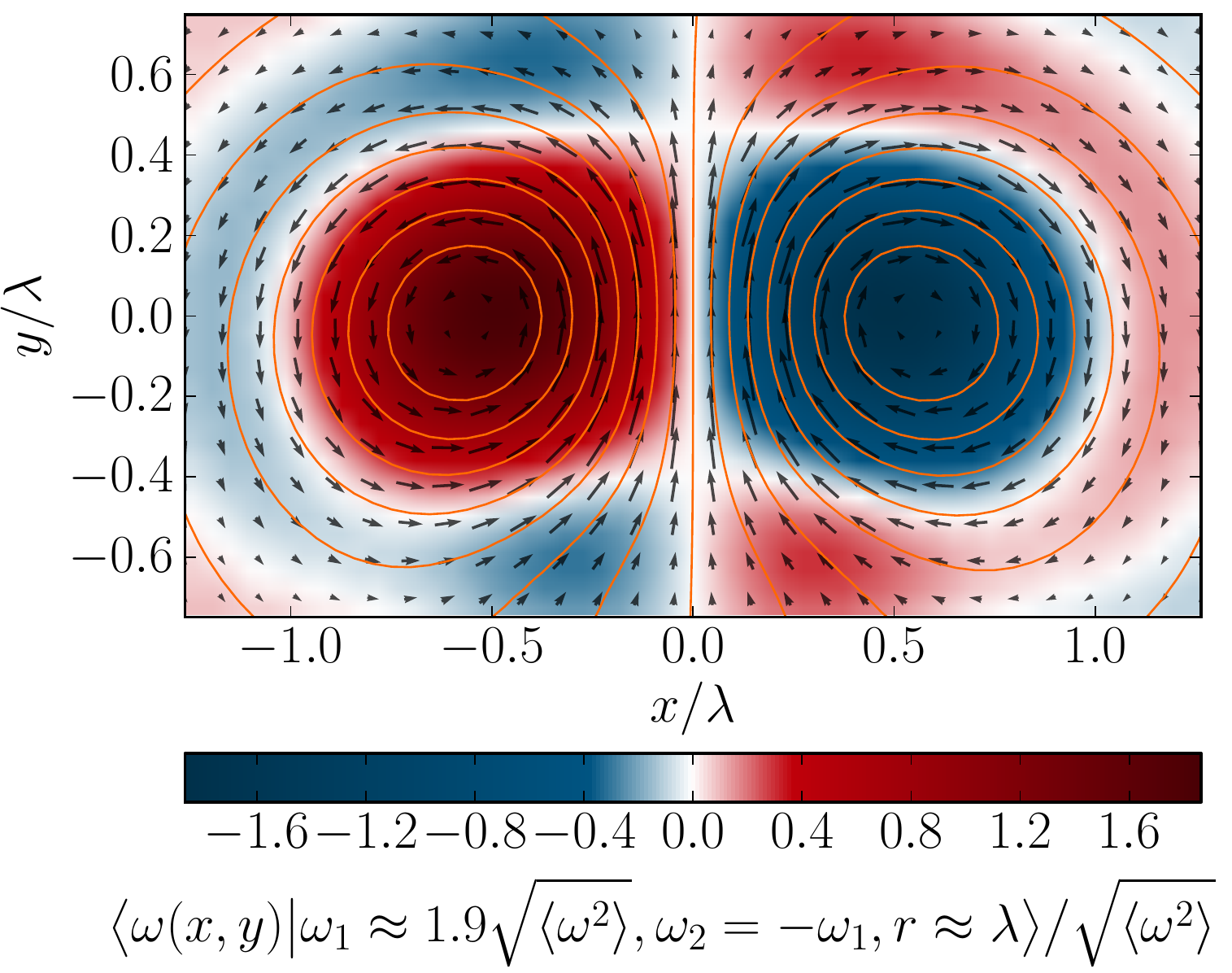}
 \includegraphics[width=0.5\textwidth]{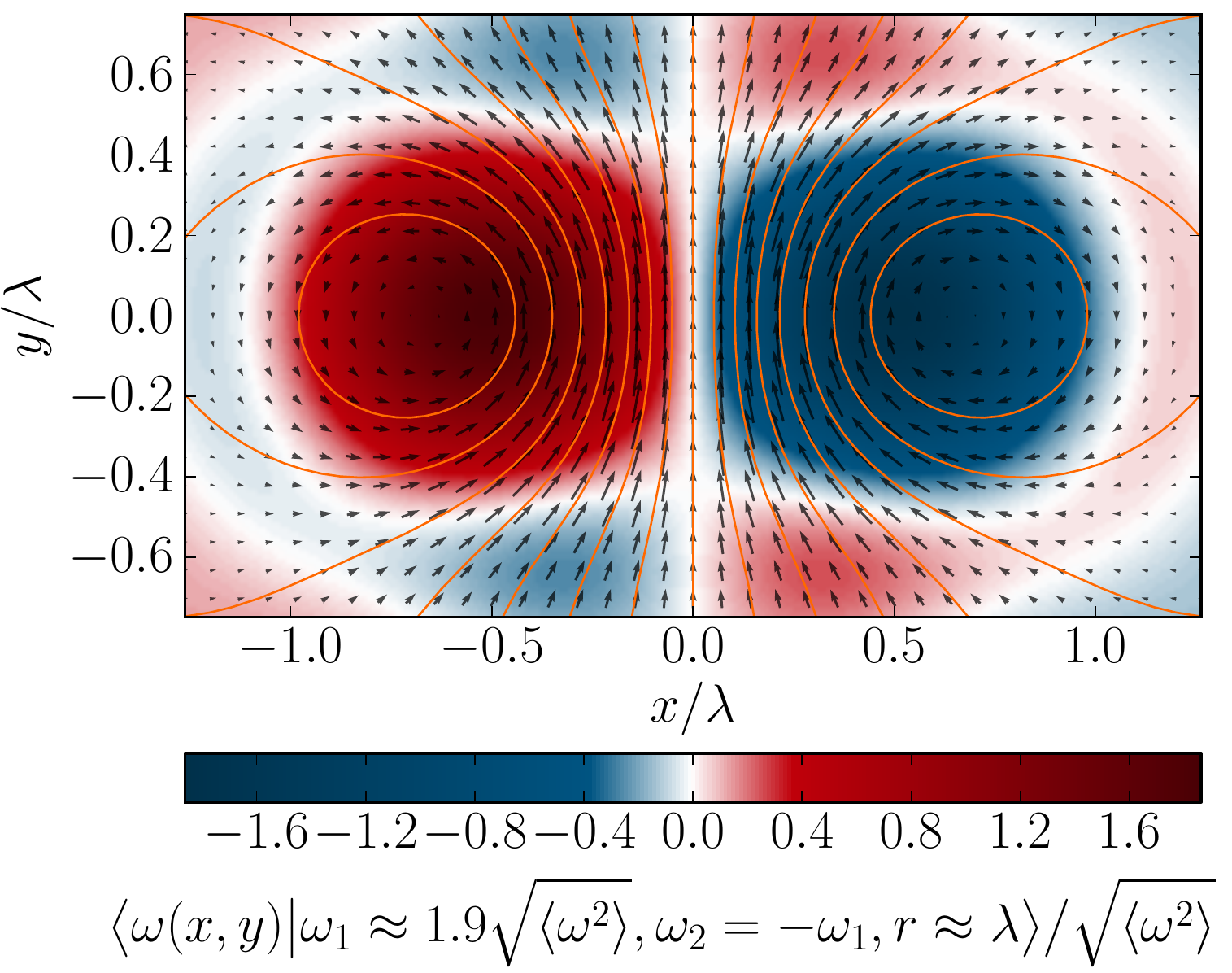}
 \caption{Conditional vorticity field for two fixed vorticities in two-dimensional turbulence in the inverse cascade regime. The vorticity field is localized near the two base points and shows similarities to the vorticity field of two vortex patches. The induced velocity field is marked with vectors and streamlines. Left: Measured directly from DNS. Right: Re-constructed following the Gaussian approximation in eq.~\eqref{eq:gauss_approx}.}
 \label{fig:2d_2p_vorticity_conaver}
\end{figure}

In this section we review results of an analysis of two-dimensional turbulence based on the kinetic equation approach and the results derived in section~\ref{lmn_2d}.
The performed analysis along the lines of the LMN hierarchy is based on a DNS of two-dimensional turbulence in the inverse cascade regime.
The DNS data comes from a pseudospectral simulation of the two-dimensional Navier-Stokes equation in vorticity formulation (cf.~fig.~\ref{fig:2d_vorticity_viz}).
Forcing at small scales generates an inverse cascade, where energy is transported to large-scale motions, which are destroyed by large-scale dissipation in order to generate stationary statistics.
These simulations show a constant energy flux within the inertial range.
Details on the setup of the simulations can be found in \cite{Friedrich2010arx}.
In the following we present the conditional averages appearing at the second level of the LMN hierarchy (obtained from DNS) and compare them with a Gaussian approximation.

\subsection{Two-Point Kinetic Equation}
The properties of the vorticity field in the inverse cascade has been analyzed on the basis of the two-point statistics.
Considering eq.~\eqref{eq:Npointwithconave} for the case of $N=2$ and additionally assuming stationary statistics, we obtain
\begin{align}
  \bigl[\nabla_{{\bm x}_1} \cdot &\langle {\bm u}({\bm x}_1,t)|\omega_1,{\bm x}_1,\omega_2,{\bm x}_2\rangle +\nabla_{{\bm x}_2}\cdot \langle {\bm u}({\bm x}_2,t)|\omega_1,{\bm x}_1,\omega_2,{\bm x}_2\rangle\bigr] f(\omega_1,{\bm x}_1,\omega_2,{\bm x}_2) \nonumber \label{eq:2ppdfeq}  \\
=-\Bigl[ &\frac{\partial }{\partial \omega_1}\langle \mu({\bm x}_1,t)|\omega_1,{\bm x}_1,\omega_2,{\bm x}_2\rangle +\nonumber\\
  &\frac{\partial }{\partial \omega_2}\langle \mu({\bm x}_2,t)|\omega_1,{\bm x}_1,\omega_2,{\bm x}_2\rangle \Bigr]f(\omega_1,{\bm x}_1,\omega_2,{\bm x}_2)\eqspace.
\end{align}
As has been shown in \cite{Friedrich2010arx}, the unclosed terms are specified by the conditional vorticity field $\langle \omega({\bm x},t)|\omega_1,{\bm x}_1,\omega_2,{\bm x}_2 \rangle,$ which especially allows us to evaluate the conditional velocity field $\langle {\bm u}({\bm x},t)|\omega_1,{\bm x}_1,\omega_2,{\bm x}_2 \rangle$.
In \cite{Friedrich2010arx} these two terms have been determined from DNS data.
The result is shown in fig.~\ref{fig:2d_2p_vorticity_conaver}.
The fields reveal close similarities with fields generated by a vorticity field which is strongly localized at the base points ${\bm x}_1$, ${\bm x}_2$.
Therefore, the characteristic equations associated to the kinetic equations for few points are closely related to the dynamics of localized vortex patches, which yields a direct connection between the statistical description of two-dimensional turbulence and vorticity dynamics.
In the simplest case these patches can be approximated by point vortices.
Hence, the knowledge about the dynamics of point vortex systems (we refer the reader to the monograph of Newton \cite{Newton2001book} and the review articles by Aref \cite{Aref2007jmp}) turns out to be of great importance.
Furthermore, the conditional dissipation and forcing field turns out to be a linear function in $\omega_1$, $\omega_2$, see \cite{Friedrich2010arx}.
As a consequence one may visualize the two-point statistics on the basis of vortex models for two localized vorticity patches.

\subsection{Gaussian Approximation to the Conditional Velocity Field}
The simplicity of the above observation has motivated further analytical investigation in \cite{Friedrich2010arx}.
A first approximation to the conditional velocity field can be obtained on the basis of a Gaussian approximation of the conditional PDF.
The general expression \eqref{eq:analyticCond} evaluated for Gaussian statistics yields the following expression for the conditional vorticity field:
\begin{align}\label{eq:gauss_approx}
  \langle & \omega({\bm x},t) |\omega_1,{\bm x}_1,\omega_2,{\bm x}_2\rangle \nonumber \\ 
  & =C({\bm x}-{\bm x}_1)\left[C_{11}^{-1}\omega_1+C_{12}^{-1}\omega_2\right]
  +C({\bm x}-{\bm x}_2)\left[C_{21}^{-1}\omega_1+C_{22}^{-1}\omega_2\right]
\end{align}
Here, $C({\bm x}-{\bm x}')=\langle \omega({\bm x},t)\omega({\bm x}',t)\rangle$ is the two-point correlation function of the vorticity field (cf. eq.~\eqref{analyticCondGauss} for the definitions of the further quantities).
This revealing result shows that the conditional vorticity field in fact is made of vortex patches with vortex profiles given by the two-point correlation.

The conditional velocity field can be determined from an application of  Biot-Savart's law (cf. eq.~\eqref{eq:biot_savart}).
The evaluation of the corresponding formula leads to the introduction of a screened velocity field:
\begin{equation}
  {\bm U}_S({\bm x}-{\bm x}_i)=\int \! \di {\bm x}' \, {\bm U}({\bm x}-{\bm x}') \, C({\bm x}'-{\bm x}_i)
\end{equation}
The subscript $S$ shall indicate that the velocity field ${\bm U}_S$ may be regarded as the field of a \textit{screened} vortex similar to the notion of Landau quasi-particles.
For a point vortex, the velocity field is obtained by replacing $C({\bm x}-{\bm x}')$ with a $\delta$-function.
For widely separated base points the conditional velocity field turns out to be a superposition of the fields of two vortices localized at ${\bm x}_1$, ${\bm x}_2$:
\begin{equation}
  \langle {\bm u}({\bm x},t)|\omega_1,{\bm x}_1,\omega_2,{\bm x}_2\rangle =\frac{1}{C(\bm 0)}\left[{\bm U}_S({\bm x}-{\bm x}_1)\omega_1 +{\bm U}_S({\bm x}-{\bm x}_2)\omega_2\right]
\end{equation}
However, the velocity profile is different from the one of point vortices, due to the fact that the turbulent surrounding has been included in terms of the conditional Gaussian approximation.
This fact allows one to visualize the characteristic equations as describing the dynamics of quasi-vortices in the sense of Landau quasi-particles.
Fig.~\ref{fig:2d_2p_vorticity_conaver} compares the conditional velocity fields determined from DNS with the Gaussian approximation.

\subsection{Failure of the Gaussian Approximation}
Although the Gaussian approximation yields a reasonable representation of the conditional vorticity and velocity field it does not reproduce the energy transport of the inverse cascade.
The conditional velocity fields
\begin{align}
  \langle {\bm u}({\bm x}_1,t)|\omega_1,{\bm x}_1,\omega_2,{\bm x}_2\rangle &=  \frac{1}{C(\bm 0)}{\bm U}_S({\bm x}_1-{\bm x}_2)\omega_2 \nonumber \\
  \langle {\bm u}({\bm x}_2,t)|\omega_1,{\bm x}_1,\omega_2,{\bm x}_2\rangle &= \frac{1}{C(\bm 0)}{\bm U}_S({\bm x}_2-{\bm x}_1)\omega_1
\end{align}
only have a component in the direction ${\bm e}_z\times ({\bm x}_1-{\bm x}_2)$, which has important implications for the two-point statistics.
This can be seen by noticing that for homogeneous flows the left-hand side of eq.~\eqref{eq:2ppdfeq} can be rewritten as
\begin{align}
  &[\nabla_{{\bm x}_1} \cdot \langle {\bm u}({\bm x}_1,t)|\omega_1,{\bm x}_1,\omega_2,{\bm x}_2\rangle +\nabla_{{\bm x}_2}\cdot \langle {\bm u}({\bm x}_2,t)|\omega_1,{\bm x}_1,\omega_2,{\bm x}_2\rangle] f(\omega_1,{\bm x}_1,\omega_2,{\bm x}_2) \nonumber \\
  =&[\nabla_{{\bm r}} \cdot \langle {\bm u}({\bm x}_2,t)-{\bm u}({\bm x}_1,t) | \omega_1,{\bm x}_1,\omega_2,{\bm x}_2\rangle f(\omega_1,{\bm x}_1,\omega_2,{\bm x}_2)]
\end{align}
with $\bm r=\bm x_2 - \bm x_1$.
As a result, this convective term vanishes for the Gaussian approximation.
Calculating moments from eq.~\eqref{eq:2ppdfeq} shows that it is precisely this term which is related to the enstrophy and energy fluxes across scales, which then also implies that these fluxes vanish in a Gaussian approximation.
For a more detailed discussion of the Gaussian approximation we refer the reader to \cite{Friedrich2010arx}.

A particularly intuitive interpretation of this statistical result is achieved with the help of the method of characteristics, which evolve according to
\begin{align}
  \dot {\bm x}_1 &= \frac{\omega_2}{C(\bm 0)} {\bm U}_S({\bm x}_1-{\bm x}_2) \nonumber \\
  \dot {\bm x}_2 &= \frac{\omega_1}{C(\bm 0)} {\bm U}_S({\bm x}_2-{\bm x}_1) \nonumber \\
  \dot \omega_1 &= \langle \mu({\bm x}_1,t)|\omega_1,{\bm x}_1,\omega_2,{\bm x}_2\rangle \nonumber \\
  \dot \omega_2 &= \langle \mu({\bm x}_2,t)|\omega_1,{\bm x}_1,\omega_2,{\bm x}_2\rangle \eqspace .
\end{align}
This set equations shows that statistically an azimuthal motion with time-dependent vorticity amplitudes is induced, a change of distance however is absent.

What apparently is missing in the Gaussian approximation is a relative motion of the two quasi-vortices, i.\,e.~a motion in the direction of  ${\bm r}$.
Therefore, a straightforward generalization of the characteristic equations takes the form
\begin{align}\label {char}
  \dot {\bm x}_1 &= \frac{\omega_2}{C(0)} {\bm U}({\bm x}_1-{\bm x}_2) +\frac{{\bm x}_1-{\bm x}_2}{|{\bm x}_1-{\bm x}_2|}H(|{\bm x}_1-{\bm x}_2|, \omega_1,\omega_2) \nonumber \\
  \dot {\bm x}_2 &= \frac{\omega_1}{C(0)} {\bm U}({\bm x}_2-{\bm x}_1) +\frac{{\bm x}_2-{\bm x}_1}{|{\bm x}_2-{\bm x}_1|}H(|{\bm x}_2-{\bm x}_1|, \omega_2,\omega_1) \eqspace ,
\end{align}
with a function $H$ that depends on distance only.
This relative motion is the signature of the inverse cascade in the characteristic equations associated to the kinetic two-point equation: The surrounding fluid turbulence leads to an effective attraction or repulsion of the quasi-vortices in the characteristic equations \eqref{char}.

A successful theory of the inverse cascade has to provide an answer to the question on the dynamical origin of the relative motion of two quasi-vortices.
Recently, it has been suggested that the relative motion is induced by an elliptical deformation of the vorticity profile of the vortices, due to the mechanism of vortex thinning, i.\,e.~an elongation and thinning of the quasi-vortices due to its mutual interaction.
Modeling these effects on the basis of a generalization of Onsager's point vortex models by introducing
elliptical point vortices which can respond to the large-scale field generated by other vortices has been shown to be successful.
The corresponding dynamics exhibits the existence of an inverse cascade by the formation of clusters of vortices with the same circulations \cite{Friedrich2011arx}.

\section{Velocity Statistics in Three-Dimensional Turbulence}
\subsection{LMN Hierarchy for the Velocity}
The hierarchy of evolution equations for the $N$-point probability
distributions of the velocity field 
\begin{equation}
  f(\lbrace\bm{u}_{l},\bm{x}_{l}\rbrace,t)=\left \langle \prod_{l=1}^N \delta\bigl(\bm{u}_{l}-\bm{u}(\bm{x}_{l},t)\bigr) \right \rangle
\end{equation}
has been formulated independently by Monin and Lundgren \cite{lundgren67pof,monin67pmm}.
The starting point for the derivation is the Navier-Stokes equation for an infinitely extended incompressible fluid
\begin{align}
  &\left(\frac{\partial}{\partial t} +\bm{u}(\bm{x},t)\cdot\nabla\right)\bm{u}(\bm{x},t) \nonumber \\ &=-\nabla p+\nu\Delta\bm{u}(\bm{x},t)+\bm{F}(\bm{x},t)  \nonumber \\
  &=  -\int \! \di \bm{x}' \, \bm{K}(\bm{x}-\bm{x}')\nabla_{\bm{x}'}\cdot\left[\bm{u}(\bm{x}',t)\cdot\nabla_{\bm{x}'}\bm{u}(\bm{x}',t)\right]+\nu\Delta\bm{u}(\bm{x},t)+\bm{F}(\bm{x},t)\eqspace,\label{eq:NS-nonlocal}
\end{align}
where the pressure term has been explicitly represented as a functional
of the velocity field using the kernel
\begin{equation}
  \bm{K}(\bm{x}-\bm{x}')=\frac{1}{4\pi}\frac{\bm{x}-\bm{x}'}{\left|\bm{x}-\bm{x}'\right|^3}\eqspace.\label{eq:kernel}
\end{equation}
The field $\bm{F}(\bm{x},t)$ is an external force which drives the fluid motion.
For the sake of simplicity we neglect the forcing in this section.
The evolution equation for the $N$-point velocity PDF can be obtained with the method of fine-grained PDFs as discussed before, resulting in
\begin{align}
  \biggl(\frac{\partial}{\partial t}+&\sum_{j=1}^{N}\bm{u}_{j}\cdot\nabla_{\bm{x}_{j}}\biggr)f\left(\left\{ \bm{u}_{l},\bm{x}_{l}\right\} ,t\right)= \nonumber \\
  &\sum_{j=1}^{N}\nabla_{\bm{u}_{j}}\cdot\int \! \di \bm{u}'\di\bm{x}' \, \bm{K}(\bm{x}_{j}-\bm{x}')\left(\bm{u}'\cdot\nabla_{\bm{x}'}\right)^{2}f(\bm{u}',\bm{x}',\lbrace\bm{u}_{l},\bm{x}_{l}\rbrace,t) \nonumber \\
  -\nu&\sum_{j=1}^{N}\nabla_{\bm{u}_{j}}\cdot\int \! \di \bm{u}'\di \bm{x}'\,\delta\bigl(\bm{x}_{j}-\bm{x}'\bigr)\bm{u}'\Delta_{\bm{x}'}f(\bm{u}',\bm{x}',\lbrace\bm{u}_{l},\bm{x}_{l}\rbrace,t)\eqspace.\label{eq:LMN-3d}
\end{align}
A readable discussion of the steps involved in the derivation can be found in ref.~\cite{lundgren67pof}.
This equation couples to the $(N+1)$-point PDF through the two terms on the right-hand side, which represent the contributions due to the pressure gradient and the Laplacian of the velocity field.
Therefore what we just obtained is an infinite hierarchy of coupled evolution equations, the LMN hierarchy for the velocity.

As a side note, we want to mention that the Friedmann-Keller equation for the correlation functions $\langle u_{i}(\bm{x}_{1},t)\dots u_{j}(\bm{x}_{N},t)\rangle$ can be obtained from the LMN hierarchy by representing the correlation functions as moments of the corresponding PDFs and using the evolution equation \eqref{eq:LMN-3d}.

As discussed before, it is possible to truncate the hierarchy at any level through the introduction of conditional averages for the pressure gradient and the dissipation.
Doing this at the $N$-th level we obtain
\begin{align}
  \biggl(\frac{\partial}{\partial t}+&\sum_{j=1}^{N}\bm{u}_{j}\cdot\nabla_{\bm{x}_{j}}\biggr) f\left(\left\{ \bm{u}_{l},\bm{x}_{l}\right\} ,t\right) = \nonumber \\
  &-\sum_{j=1}^{N}\nabla_{\bm{u}_{j}}\cdot\langle-\nabla p(\bm{x}_{j},t)|\lbrace\bm{u}_{l},\bm{x}_{l}\rbrace\rangle f(\lbrace\bm{u}_{l},\bm{x}_{l}\rbrace,t) \nonumber \\
  &-\sum_{j=1}^{N}\nabla_{\bm{u}_{j}}\cdot\langle\nu\Delta_{\bm{x}_{j}}\bm{u}(\bm{x}_{j},t)|\lbrace\bm{u}_{l},\bm{x}_{l}\rbrace\rangle f(\lbrace\bm{u}_{l},\bm{x}_{l}\rbrace,t)\nonumber \\
  &-\sum_{j=1}^{N}\nabla_{\bm{u}_{j}}\cdot\langle\bm{F}(\bm{x}_{j},t)|\lbrace\bm{u}_{l},\bm{x}_{l}\rbrace\rangle f(\lbrace\bm{u}_{l},\bm{x}_{l}\rbrace,t)\eqspace,
\end{align}
where we have now included the forcing.
This evolution equation is formally closed but contains the unknown conditional averages of the pressure gradient and the dissipation.
These terms can be either modeled or estimated from experiments or simulations.

\subsection{Single-Point Velocity Statistics} \label{sec:singlepointvelocity}
We now come to a closer investigation of the single-point velocity PDF in homogeneous isotropic turbulence.
Although this is maybe one of the simplest statistical quantities to consider, there has been an ongoing discussion in the past whether this quantity exhibits deviations from Gaussianity.
Gaussianity is often assumed based on an argumentation exploiting the central limit theorem.
Numerical and experimental investigations, however, give evidence for slight deviations from Gaussianity, especially indicating the occurrence of sub-Gaussian tails \cite{vincent91jfm,noullez97jfm,gotoh02pof}.
From the theory side, varying opinions can be found in the literature.
Falkovich and Lebedev have used the instanton formalism to argue in favor of sub-Gaussian tails depending on the external forcing \cite{falkovich97prl}.
Gaussian PDFs have been found for decaying turbulence by Ulinich and Lyubimov \cite{ulinich69spj} and later by Hosokawa \cite{hosokawa08pre} or in the case of the cross-independence hypothesis by Tatsumi and coworkers \cite{tatsumi04fdr}.

By combining the kinetic equation of the single-point velocity with DNS results, we will argue here for the existence of deviations from Gaussianity caused by statistical correlations, e.\,g.~of the dissipation field and the velocity field.
For more details we kindly refer the reader to refs.~\cite{wilczek11jfm,wilczek11epl,wilczek11phd}.

From now on we will drop the argument $(\bm x,t)$ for all realizations.
We remind the reader, however, that the first argument of a conditional average is understood as a realization whereas the condition is specified by the corresponding sample-space variable.

Introducing conditional averages on the level of the first equation of the hierarchy we arrive at
\begin{equation}\label{eq:f1kinetic}
  \frac{\partial}{\partial t} f(\bm u,\bm x, t) + \bm u \cdot \nabla f(\bm u,\bm x,t) = -\nabla_{\bm u} \cdot \Bigl[ \big\langle -\nabla p + \nu \Delta \bm u + \bm F \big | \bm u \big\rangle \, f(\bm u,\bm x,t) \Bigr] \eqspace .
\end{equation}
The left-hand side is closed, however, the right-hand side contains unclosed terms involving the joint statistics of the pressure gradient, viscous diffusion and the external forcing with the velocity.

To make a connection to DNS results, this equation has to be simplified with the help of statistical symmetries.
As exemplified above, the PDF is independent of the spatial coordinate for homogeneous flows.
As a result, the advective term on the left-hand side vanishes.
Homogeneity further lets us derive the relation
\begin{equation}\label{eq:homrel}
  \frac{\partial^2}{\partial x_k^2} f(\bm u, t) = 0 =-\frac{\partial }{\partial u_i}  \bigg \langle  \frac{\partial^2 u_i}{\partial x_k^2}  \bigg | \bm u  \bigg \rangle f(\bm u, t) +\frac{\partial }{\partial u_i}\frac{\partial }{\partial u_j}  \bigg \langle \frac{\partial u_i}{\partial x_k} \frac{\partial u_j}{\partial x_k} \bigg | \bm u  \bigg \rangle f(\bm u, t) \eqspace ,
\end{equation}
which is a multi-dimensional generalization of the relations introduced and investigated by Ching and Pope \cite{pope1993pof,ching93prl,ching96pre}.
Inserting this result into the kinetic equation~\eqref{eq:f1kinetic} leads to
\begin{equation}\label{eq:f1homogeneity}
  \frac{\partial}{\partial t}f(\bm u, t)=-\frac{\partial}{\partial u_i} \bigg \langle -\frac{\partial}{\partial x_i} p +  F_i  \bigg | \bm u  \bigg \rangle f(\bm u, t) - \frac{\partial }{\partial u_i}\frac{\partial }{\partial u_j} \bigg \langle \nu \frac{\partial u_i}{\partial x_k} \frac{\partial u_j}{\partial x_k}  \bigg | \bm u  \bigg \rangle f(\bm u, t) \eqspace .
\end{equation}
By this we have replaced the conditional diffusive term with the conditional (pseudo-)dissipation tensor
\begin{equation}
  D_{ij}(\bm u)= \bigg \langle  \nu \frac{\partial u_i}{\partial x_k} \frac{\partial u_j}{\partial x_k} \bigg | \bm u  \bigg \rangle \eqspace .
\end{equation}
For isotropic turbulence, all of the appearing quantities have to be invariant with respect to rotations.
As a consequence, the velocity PDF depends on the velocity magnitude only and is determined by the PDF of the velocity magnitude by the relation
\begin{equation}
  \tilde f(u,t) = 4\pi u^2 f(\bm u,t)  \eqspace .
\end{equation}

The conditional vectors appearing in \eqref{eq:f1kinetic} take the form
\begin{subequations}
\begin{align}
  \left\langle -\nabla p \big | \bm u \right\rangle  = \Pi(u) \hat{\bm u} 
  &\qquad \Pi(u)=\left\langle - \hat{\bm u} \cdot \nabla p \big | u \right\rangle \label{eq:iso-pressure}\\ 
  \left\langle \nu \Delta \bm u \big | \bm u \right\rangle = \Lambda(u) \hat{\bm u}
  &\qquad \Lambda(u)=\left\langle \nu  \hat{\bm u} \cdot \Delta \bm u \big | u \right\rangle \label{eq:iso-laplace}\\
  \left\langle \bm F \big | \bm u \right\rangle = \Phi(u) \hat{\bm u} 
  &\qquad \Phi(u)=\left\langle \hat{\bm u} \cdot \bm F \big | u \right\rangle \label{eq:iso-force} \eqspace .
\end{align}
\end{subequations}
Note that the conditional averages $\Pi$, $\Lambda$ and $\Phi$ are scalar functions of a scalar argument, which makes an estimation from DNS data feasible.
The conditional dissipation tensor can be simplified in a similar manner and can be written down in terms of its eigenvalues according to
\begin{subequations}
\begin{align}
  D_{ij}(\bm u) &= \mu(u) \, \delta_{ij} + \left[ \lambda(u)-\mu(u)\right]  \frac{u_i u_j}{u^2}\label{eq:D-lambda-mu}\\
  \mu(u) &= \frac{1}{4}\left\langle  \varepsilon+\nu \omega^2 \big | u\right\rangle-\frac{1}{2}\left\langle \nu(\mathrm{A}^T\hat{\bm u})^2\big | u\right\rangle \label{eq:mu_relation} \\ 
  \lambda(u) &= \left\langle \nu(\mathrm{A}^T\hat{\bm u})^2\big | u\right\rangle \label{eq:lambda_relation} \eqspace .
\end{align}
\end{subequations}
Here, the conditionally averaged kinetic energy dissipation, the squared vorticity and the transpose of the velocity gradient tensor $A_{ij}=\frac{\partial u_i}{\partial x_j}$ enter, clarifying that the single-point velocity statistics depends on the correlation of the dissipation field and the velocity field.
Taking all of these results together, eqs.~\eqref{eq:f1kinetic} and \eqref{eq:f1homogeneity} can be recast in the forms
\begin{subequations}
\begin{align}
  \frac{\partial}{\partial t} \tilde f(u,t)&=-\frac{\partial}{\partial u} \bigl( \Pi(u,t)+\Lambda(u,t)+\Phi(u,t) \bigr) \tilde f(u,t)\label{eq:pdfveliso}\\
  \frac{\partial}{\partial t} \tilde f(u,t)&=-\frac{\partial}{\partial u} \left( \Pi(u,t)+\Phi(u,t)-\frac{2\mu(u,t)}{u} \right) \tilde f(u,t) - \frac{\partial^2}{\partial u^2} \lambda(u,t) \tilde f(u,t)\label{eq:pdfvelhomoiso} \eqspace .
\end{align}
\end{subequations}

Two interesting conclusions can be drawn for stationary flows.
First, eq.~\eqref{eq:pdfveliso} implies 
\begin{equation}
  \Pi(u)+\Lambda(u)+\Phi(u)=0 \eqspace ,
\end{equation}
i.\,e.~balance of the pressure term, the diffusive term and the forcing term for each fixed value of velocity.
Even more importantly, a stationary solution can be derived for \eqref{eq:pdfvelhomoiso} taking the form
\begin{equation}\label{eq:statsol}
  \tilde f(u)=\frac{{\cal N}}{\lambda(u)} \exp \int_{u_0}^u \! \mathrm{d}u' \, \frac{ -\Pi(u')-\Phi(u')+\frac{2}{u'}\mu(u')}{\lambda(u')} \eqspace ,
\end{equation}
where ${\cal N}$ is a normalization constant depending on $u_0$.
This solution is an exact, yet unclosed relation.
Without further input of numerical data, the conclusion to be drawn from these theoretical results is that the shape of the PDF is determined by the conditional averages which encode the correlations of the various turbulent fields appearing in the Navier-Stokes equation with the velocity field.

\begin{figure}[t]
 \includegraphics[width=0.5\textwidth]{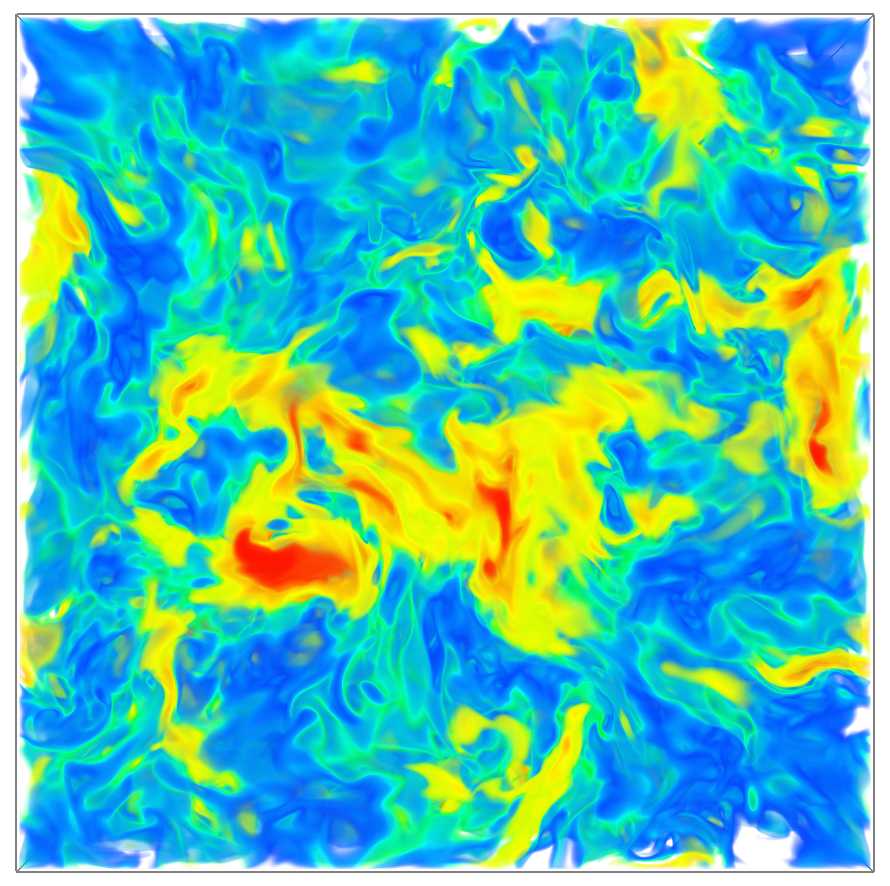}
 \includegraphics[width=0.5\textwidth]{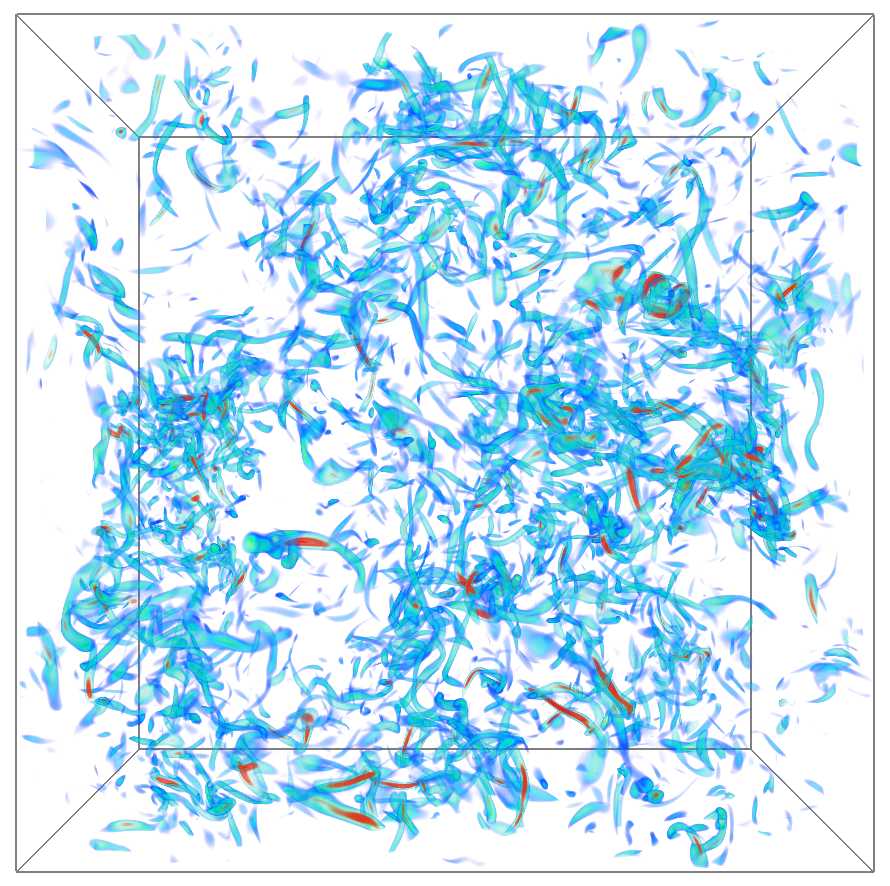}
 \caption{Visualization of the velocity (left) and vorticity field (right) in three-dimensional homogeneous isotropic turbulence. The figures show the color-coded magnitude of the full vector, where blue represents low and red high values. Additional visualizations to be found at \cite{url:turbulenceteamms}.}
 \label{fig:3d_viz}
\end{figure}

Before turning to the DNS data, we would like to demonstrate how a ``naive'', yet rational closure can be established by simple physical arguments \cite{wilczek11epl}.
To this end we make use of the fact that the conditional averages can be reduced to ordinary averages by integration, which lets us introduce a number of integral constraints.
For example, for homogeneous isotropic flow, the pressure term has to vanish, implying
\begin{equation}
  0 = \left \langle \bm{u}\cdot\nabla p \right \rangle = \int_{0}^{\infty} \! \mathrm{d}u \, \left \langle \bm{u}\cdot\nabla p  \big | u \right \rangle \tilde f(u) = -\int_{0}^{\infty} \!  \mathrm{d}u \, u\,\Pi(u) \tilde f(u) \eqspace .
  \label{eq:pressure-int-constraint}
\end{equation}
The dissipative terms have to yield the mean kinetic energy dissipation
\begin{subequations}
\begin{align}
  -\left\langle \varepsilon \right\rangle &= \left\langle \nu \bm u \cdot \Delta \bm u\right\rangle=  \int_{0}^{\infty} \! \mathrm{d}u \, u\,\Lambda(u) \tilde f(u) \label{eq:Lambda-int-constraint} \\
  \left\langle \varepsilon \right\rangle &= \left\langle \mathrm{Tr}(\mathrm{D}) \right\rangle=  \int_{0}^{\infty} \! \mathrm{d}u \, [\lambda(u)+2\mu(u)] \tilde f(u) \label{eq:lambda-mu-int-constraint} \eqspace .
\end{align}
\end{subequations}
As for stationary flows the forcing terms balances the mean energy dissipation, we obtain
\begin{equation}
  \left\langle \varepsilon \right\rangle = \left\langle \bm u \cdot \bm F\right\rangle=  \int_{0}^{\infty} \! \mathrm{d}u \, u\,\Phi(u) \tilde f(u) \eqspace .
  \label{eq:Phi-int-constraint}
\end{equation}
Based on these constraints, a lowest-order approximation of the conditional averages yields
\begin{subequations}
\begin{align}
  \Pi_0(u) &= 0 \label{eq:decoupling1} \\
  \Phi_0(u) &= \frac{\langle\varepsilon\rangle}{3\sigma^2}u \label{eq:decoupling2} \\
  \Lambda_0(u) &= -\frac{\langle\varepsilon\rangle}{3\sigma^2}u \label{eq:decoupling3} \eqspace ,
\end{align}
\end{subequations}
where $\sigma=\sqrt{\langle \bm u^2/3 \rangle}$.
In this approximation, pressure effects are neglected completely and the forcing and diffusive term are linear functions of the sample-space velocity.
For the conditional dissipation tensor we arrive at
\begin{equation}\label{eq:decoupling4}
  \lambda_0(u)=\mu_0(u)=\frac{\langle \varepsilon \rangle}{3} \eqspace ,
\end{equation}
i.\,e., the two eigenvalues are identical and can be expressed in terms of the average rate of kinetic energy dissipation.
This last approximation can also be understood in terms of a simple decoupling argument: Assuming a scale-separation between the small-scale dissipation field and the comparably large-scale velocity field (see fig.~\ref{fig:3d_viz} for a visualization of the velocity and the vorticity field) implies that the conditional averages reduce to ordinary averages.
Evaluating the stationary solution \eqref{eq:statsol} with these approximation yields a Maxwellian distribution of the velocity magnitude
\begin{equation}
  \tilde{f}(u)=\sqrt{\frac{2}{\pi}}\frac{u^2}{\sigma^3}\exp\left( -\frac{u^2}{2\sigma^2} \right) \eqspace ,
  \label{eq:maxwellian}
\end{equation}
which corresponds to a Gaussian distribution of the velocity vector,
\begin{equation}
  f(\bm u)=\frac{1}{(2\pi\sigma^{2})^{3/2}}\exp\left( -\frac{\bm u^2}{2\sigma^2} \right) \eqspace .
  \label{eq:gaussian}
\end{equation}
This interesting result shows that a Gaussian single-point velocity PDF can be obtained as a lowest-order approximation.
The DNS results in the following, however, will show that pronounced deviations from this approximation exist and in combination lead to deviations from Gaussianity.

\begin{figure}[t]
 \includegraphics[width=0.5\textwidth]{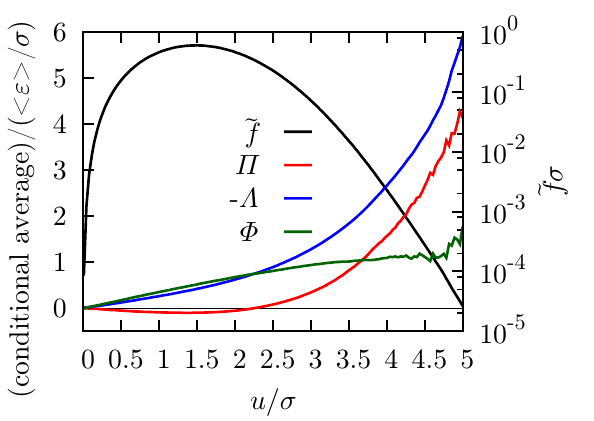}
 \includegraphics[width=0.5\textwidth]{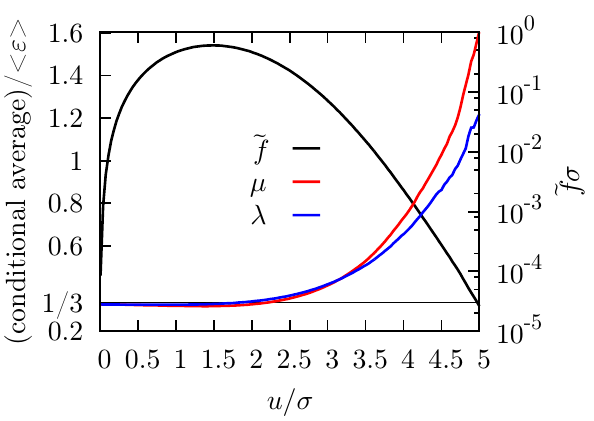}
 \caption{Conditional averages arising in the kinetic equation for the single-point velocity PDF.}
 \label{fig:1p_velocity_conave}
\end{figure}

The DNS results have been obtained with a standard pseudospectral simulation of the vorticity equation with periodic boundary conditions.
The forcing is acting on large scales and rescales the amplitudes of the Fourier coefficients in a wavenumber band such that the total energy of the flow is conserved while letting their phases evolve freely.
The simulation exhibits a Taylor-based Reynolds number of $R_{\lambda} \approx 112$ at a resolution of $512^3$ grid points.
To obtain good ensemble statistics, the simulation was run for more than $150$ large-eddy turnover times and two hundred snapshots of the fields were taken.
From this ensemble, the conditional averages have been estimated.
Fig.~\ref{fig:1p_velocity_conave} (left) shows the conditional averages appearing in \eqref{eq:pdfveliso}.
As can be seen from the figure, the conditional average related to the forcing term is approximately linear and thus similar to the naive closure assumption \eqref{eq:decoupling2}.
The conditional averages related to the diffusive term and the pressure gradient, however, differ strongly from the naive closure.
The conditional diffusive term is negatively correlated with the velocity, indicating that this term tends to decelerate a fluid particle on average.
The pressure term shows an interesting zero-crossing, which means that for low velocities a fluid particle is on average decelerated whereas for high velocities it is accelerated due to this term.
For a non-vanishing pressure contribution, the occurrence of a zero-crossing is necessary to fulfill the integral constraint \eqref{eq:pressure-int-constraint}.
The terms of the conditional dissipation tensor are depicted in fig.~\ref{fig:1p_velocity_conave} (right).
It becomes clear that the eigenvalues of the conditional dissipation tensor approximately coincide, i.\,e.~the conditional dissipation tensor is approximately isotropic.
For low values of velocity the naive closure assumption \eqref{eq:decoupling4} is acceptable; strong deviations, however, occur for high values of velocity indicating that high velocity regions in the flow are on average related to regions of increased kinetic energy dissipation.
As a result, a simple decoupling argument of large-scale and small-scale fields yields only a rough approximation to the observed DNS data.

\begin{figure}[t]
 \centering{\includegraphics[width=0.7\textwidth]{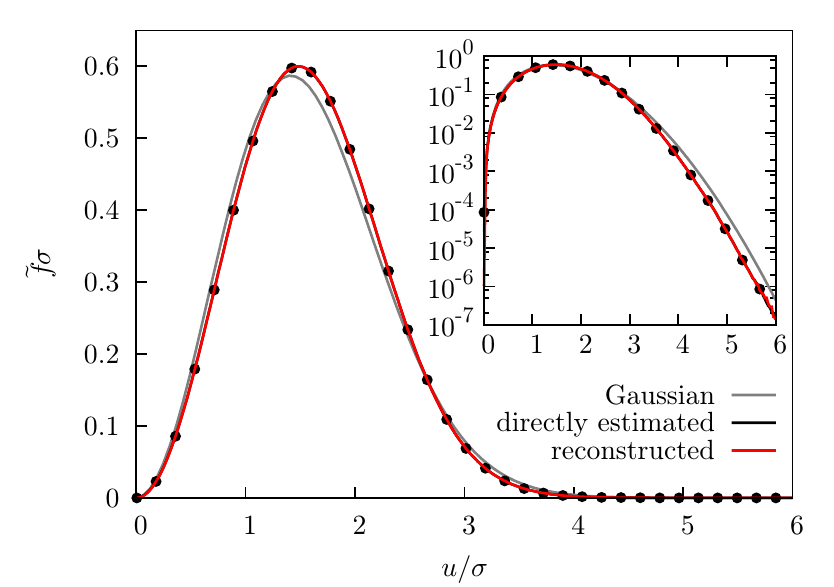}}
 \caption{Comparison of the directly estimated PDF and the one reconstructed using relation \eqref{eq:statsol}. The directly estimated PDF is marked by black dots to indicate its presence.}
 \label{fig:1p_velocity_pdf}
\end{figure}

Fig.~\ref{fig:1p_velocity_pdf} shows the evaluation of the stationary solution \eqref{eq:statsol} with the conditional averages estimated from the DNS data.
As expected, the agreement between the directly estimated PDF and the result of \eqref{eq:statsol} is perfect, but both PDFs deviate notably from the Maxwellian shape expected for a Gaussian distributed velocity vector.
Thus the conclusion can be drawn that the observed deviations from Gaussianity of the single-point velocity PDF are the result of a rather intricate interplay of statistical correlations.

\section{Vorticity Statistics in Three-Dimensional Turbulence}
\subsection{LMN Hierarchy for the Vorticity}
Analogous to the LMN hierarchy for the velocity one can derive a hierarchy of evolution equations for the vorticity PDFs, which first has been formulated by Novikov \cite{novikov68sdp}.
The hierarchy is based on the three-dimensional vorticity equation
\begin{equation}
  \left(\frac{\partial}{\partial t}+\bm{u}(\bm{x},t)\cdot\nabla\right)\bm{\omega}(\bm{x},t)={\rm S}\left(\bm{x},t\right)\bm{\omega}(\bm{x},t)+\nu\Delta\bm{\omega}(\bm{x},t)+\nabla\times\bm{F}\left(\bm{x},t\right)
\end{equation}
where ${\rm S}=\frac{1}{2}[\nabla\bm{u}+\left(\nabla\bm{u}\right)^{T}]$ is the rate-of-strain tensor.
The velocity field can be expressed as 
\begin{equation}
  \bm{u}(\bm{x},t)=\int \! \di \bm{x}' \,\bm{\omega}\left(\bm{x}',t\right)\times\bm{K}\left(\bm{x}-\bm{x}'\right)
\end{equation}
using the kernel \eqref{eq:kernel}.
In the vorticity equation the velocity and the rate-of-strain tensor represent the nonlocal character of the underlying dynamics whereas in the Navier-Stokes equation it is represented by the pressure gradient.

In the same manner as before the evolution equation for the $N$-point PDF can be derived from the dynamic equation for the corresponding quantity.
In the case of the vorticity we obtain
\begin{align}
  \frac{\partial}{\partial t}f( & \lbrace\bm{\omega}_{l},\bm{x}_{l}\rbrace,t)+\sum_{j=1}^{N}\nabla_{\bm{x}_{j}}\cdot\int \! \di \bm{x}' \di \bm{\omega}' \, \bm{\omega}'\times\bm{K}(\bm{x}_{j}-\bm{x}')f(\bm{\omega}',\bm{x}',\lbrace\bm{\omega}_{l},\bm{x}_{l}\rbrace,t)= \nonumber \\
  &-\sum_{j=1}^{N}\nabla_{\bm{\omega}_{j}}\cdot\int \! \di \bm{x}'\di \bm{\omega}' \, [\bm{\omega}'\times\bm{K}(\bm{x}_{j}-\bm{x}')]\bm{\omega}_{j}\cdot\nabla_{\bm{x}'}f(\bm{\omega}',\bm{x}',\lbrace\bm{\omega}_{l},\bm{x}_{l}\rbrace,t)\nonumber \\
  &-\sum_{j=1}^{N}\nabla_{\bm{\omega}_{j}}\cdot\int \! \di \bm{x}' \di \bm{\omega}'\, \bm{\omega}'\delta\bigl(\bm{x}_{j}-\bm{x}'\bigr)\Delta_{\bm{x}'}f(\bm{\omega}',\bm{x}',\lbrace\bm{\omega}_{l},\bm{x}_{l}\rbrace,t) \eqspace .\label{eq:vorticity-hierarchy}
\end{align}
The first term on the right-hand side stems from the vortex stretching term.
The forcing term was omitted for the sake of simplicity.
Again we see a coupling of the $N$-point PDF equation to the $(N+1)$-point equation and obtain therefore an infinite hierarchy of coupled evolution equations.

It is interesting to note here that the term resulting from the rate-of-strain tensor in eq.~\eqref{eq:vorticity-hierarchy} is related to the conditional first-order moment of vorticity.
This has to be contrasted with the fact that the term due to the pressure gradient in eq.~\eqref{eq:LMN-3d} is related to a conditional second-order moment.
Therefore, loosely speaking, less information is needed to close the PDF equation for the vorticity than the one for the velocity.
The conditional vorticity field has been investigated in \cite{mui96pre,novikov93jfr}.

An alternative representation of the LMN hierarchy for the vorticity is obtained when we express the terms
in \eqref{eq:vorticity-hierarchy} as conditional averages:
\begin{align}
  \frac{\partial}{\partial t}f(\lbrace\bm{\omega}_{l},\bm{x}_{l}\rbrace,t)&+\sum_{j=1}^{N}\nabla_{\bm{x}_{j}}\cdot\left\langle \left.\bm{u}\left(\bm{x}_{j},t\right)\right|\left\{ \bm{\omega}_{l},\bm{x}_{l}\right\} \right\rangle f(\lbrace\bm{\omega}_{l},\bm{x}_{l}\rbrace,t)= \nonumber \\
  &-\sum_{j=1}^{N}\nabla_{\bm{\omega}_{j}}\cdot\left\langle \left.{\rm S}\left(\bm{x}_{j},t\right)\bm{\omega}\left(\bm{x}_{j},t\right)\right|\left\{ \bm{\omega}_{l},\bm{x}_{l}\right\} \right\rangle f(\lbrace\bm{\omega}_{l},\bm{x}_{l}\rbrace,t) \nonumber \\
  &-\sum_{j=1}^{N}\nabla_{\bm{\omega}_{j}}\cdot\left\langle \left.\nu\Delta_{\bm{x}_{j}}\bm{\omega}\left(\bm{x}_{j},t\right)\right|\left\{ \bm{\omega}_{l},\bm{x}_{l}\right\} \right\rangle f(\lbrace\bm{\omega}_{l},\bm{x}_{l}\rbrace,t) \nonumber \\
  &-\sum_{j=1}^{N}\nabla_{\bm{\omega}_{j}}\cdot\left\langle \left.\nabla_{\bm{x}_{j}}\times\bm{F}\left(\bm{x}_{j},t\right)\right|\left\{ \bm{\omega}_{l},\bm{x}_{l}\right\} \right\rangle f(\lbrace\bm{\omega}_{l},\bm{x}_{l}\rbrace,t) \label{eq:vorticity-hierarchy-truncated}
\end{align}
In this way we again obtain a formally closed equation, which however contains the unknown conditional averages.
These averages can be either modeled or estimated from experiments or simulations.

Eq.~\eqref{eq:vorticity-hierarchy-truncated} shows that the statistics of the rate-of-strain tensor, the symmetric part of the velocity gradient tensor, plays an important role for the statistics of the vorticity, the anti-symmetric part.
Thus it is not surprising that the statistics of the full velocity gradient tensor has received a lot of interest lately.
We refrain here from going into the details and refer the interested reader to \cite{Meneveau2011} for an overview over a number of recent modeling approaches and to \cite{Tsinober2009book} for a discussion of various statistical properties of the velocity gradient tensor.

\subsection{Single-Point Vorticity Statistics} \label{sec:singlepointvorticity}
In section~\ref{sec:singlepointvelocity} we have demonstrated how a combination of the kinetic equation with DNS results has led to insights regarding the slight deviations from Gaussianity of the single-point velocity PDF.
Obviously this analysis is also feasible for the vorticity PDF.
However, the results are expected to differ strongly from the velocity case as the vorticity PDF is known to be far from Gaussian with an enhanced probability to encounter strong vorticities in the flow.
This observation can be seen as the statistical reflection of the occurrence of coherent vortex structures in the flow (see fig.~\ref{fig:3d_viz}) that leads to pronounced spatio-temporal correlations.

The kinetic equations for the single-point velocity PDF have been discussed comparably detailed above, and the derivation is completely analogous for the vorticity.
Hence we focus stronger on a discussion of the DNS results and a comparison of the velocity and vorticity statistics in this section.
We kindly refer the reader to \cite{wilczek09pre,wilczek11jfm} for a detailed presentation.

Introducing the conditional averages on the single-point level leads to a kinetic equation completely analogous to \eqref{eq:f1kinetic}:
\begin{equation}\label{eq:f1kineticvorticity}
  \frac{\partial}{\partial t} f(\bm \omega,\bm x,t) + \nabla \cdot \Bigl[ \big\langle \bm u \big | \bm \omega \big\rangle f(\bm \omega,\bm x,t) \Bigr]= -\nabla_{\bm \omega}\cdot \Bigl[ \big\langle \mathrm{S}\bm \omega + \nu \Delta \bm \omega + \nabla \times \bm F \big| \bm \omega \big\rangle f(\bm \omega,\bm x,t) \Bigr]
\end{equation}
Here appear the advective derivative of the PDF on the left-hand side and the conditionally averaged terms of the vorticity equation on the right-hand side, implying that the evolution of the PDF is governed by the statistics of vortex stretching, vorticity diffusion and the external forcing.
If we assume a homogeneous flow, the advective term on the left-hand side vanishes as neither the conditional velocity nor the PDF itself depend on the spatial coordinate.

As in the velocity case, we also exploit statistical isotropy.
While the diffusive term and the forcing term are treated like the corresponding terms from the Navier-Stokes equation, the conditional vortex stretching term requires special scrutiny.
As we condition on the single-point vorticity, the vorticity realization can be replaced by the sample-space vorticity and pulled out of the average, 
\begin{equation}
  \big\langle \mathrm{S} \bm \omega \big| \bm \omega \big\rangle = \big\langle \mathrm{S}\big| \bm \omega \big\rangle \bm \omega \eqspace ,
\end{equation}
such that the conditional rate-of-strain tensor has to be specified.
Here we have to take into account that this tensor is traceless.
This leads to the following general functional form of the unclosed terms:
\begin{subequations}
\begin{align}
  \big\langle S_{ij} \big| \bm \omega \big\rangle &= \frac{1}{2}\Sigma(\omega) \left( 3 \frac{\omega_i \, \omega_j}{\omega^2} - \delta_{ij} \right) &\quad \Sigma(\omega) &= \big\langle \widehat{\bm \omega} \cdot \mathrm{S}\widehat{\bm \omega}  \big | \omega \big\rangle \\
  \big\langle \nu \Delta \bm \omega \big| \bm \omega \big\rangle &= \Lambda(\omega) \, \widehat{\bm \omega} &\quad \Lambda(\omega) &= \big\langle \nu\widehat{\bm \omega}\cdot \Delta \bm \omega \big| \omega \big\rangle \\
  \big\langle \nabla \times \bm F \big| \bm \omega \big\rangle &= \Phi(\omega) \, \widehat{\bm \omega} &\quad \Phi(\omega) &= \big\langle \widehat{\bm \omega}\cdot(\nabla \times \bm F) \big| \omega \big\rangle
\end{align}
\end{subequations}
By this we have again reduced the unclosed terms to scalar quantities depending on the scalar argument $\omega$.
The conditional rate-of-strain tensor is fully specified by one of its eigenvalues $\Sigma(\omega)$.

Due to homogeneity, we may replace the conditional diffusive term with the so-called conditional enstrophy dissipation tensor, which is again characterized by its eigenvalues,
\begin{equation}
  D_{ij}(\bm \omega)= \bigg \langle  \nu \frac{\partial \omega_i}{\partial x_k} \frac{\partial \omega_j}{\partial x_k} \bigg | \bm \omega  \bigg\rangle = \mu(\omega) \, \delta_{ij} + \big[ \lambda(\omega) - \mu(\omega) \big] \frac{\omega_i \omega_j}{\omega^2} \eqspace .
\end{equation}
This tensor contains the statistical correlations of the vorticity with the vorticity gradients.
Inserting these results into the original kinetic equation \eqref{eq:f1kineticvorticity} lets us write down two versions of kinetic equations for the PDF of the vorticity magnitude,
\begin{subequations}
\begin{align}
  \frac{\partial}{\partial t} \tilde f(\omega,t)&=-\frac{\partial}{\partial \omega} \bigl( \Sigma(\omega,t)\,\omega+\Lambda(\omega,t)+\Phi(\omega,t) \bigr) \tilde f(\omega,t)\label{eq:pdfvorticityiso}\\
  \frac{\partial}{\partial t} \tilde f(\omega,t) &=-\frac{\partial}{\partial \omega} \left( \Sigma(\omega,t)\,\omega+\Phi(\omega,t)-\frac{2\mu(\omega,t)}{\omega} \right) \tilde f(\omega,t) - \frac{\partial^2}{\partial \omega^2} \lambda(\omega,t) \tilde f(\omega,t)\label{eq:pdfvorticityhomoiso} \eqspace .
\end{align}
\end{subequations}

\begin{figure}[t]
 \includegraphics[width=0.5\textwidth]{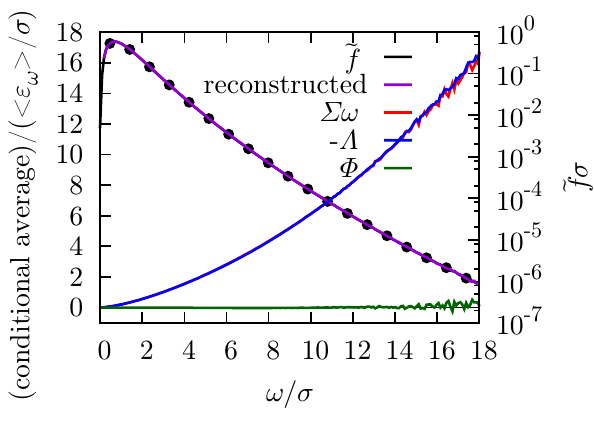}
 \includegraphics[width=0.5\textwidth]{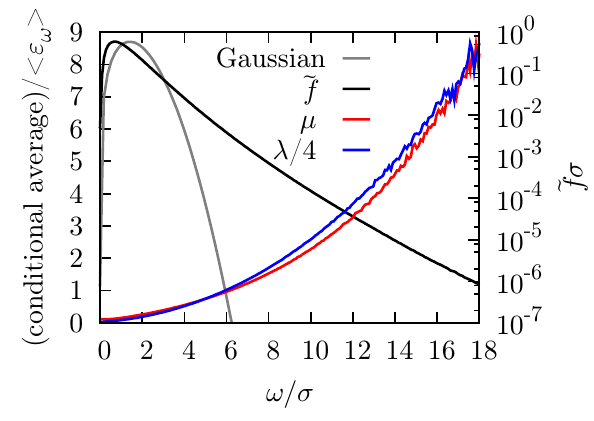}
 \caption{Conditional averages and PDF for the single-point vorticity statistics. The PDF directly estimated from data is marked by black dots to indicate its presence as it is exactly overlapped by the reconstructed PDF.}
 \label{fig:1p_vorticity_conave}
\end{figure}

For a further discussion of the stationary vorticity statistics, we turn to the DNS results, which have been obtained from the DNS data set described in section~\ref{sec:singlepointvelocity}.
The conditional averages arising in the kinetic equations are shown in fig.~\ref{fig:1p_vorticity_conave}.
In the left graph, the conditional vortex stretching term, the diffusive term and the term related to the external forcing are shown.
The vortex stretching term exhibits a nonlinear functional form indicating strong statistical correlations of the rate-of-strain-tensor and the vorticity.
As the rate-of-strain field and the vorticity field are expected to vary on comparable scales, a simple scale separation argument as discussed for the velocity does not hold.
As becomes also clear from fig.~\ref{fig:1p_vorticity_conave} (left), the conditional diffusive term almost perfectly balances the vortex stretching term.
Considering the fact derived from the kinetic equation \eqref{eq:pdfvorticityiso} that the conditional balance
\begin{equation}\label{eq:conditionalbalance}
  0=\Sigma(\omega)\,\omega+\Lambda(\omega)+\Phi(\omega)
\end{equation}
has to hold then leads to the conclusion that the conditional forcing term has to vanish approximately, which is numerically confirmed in fig.~\ref{fig:1p_vorticity_conave} (left).
We would like to mention that this result has already been theoretically predicted by Novikov \cite{novikov93jfr} on the basis of a high Reynolds number argument.
As a result, the single-point statistics of the vorticity is rather independent of the external forcing and the approximate relation
\begin{equation}
  0\approx\Sigma(\omega)\,\omega+\Lambda(\omega)
\end{equation}
holds.
In this sense, this statistical quantity purely reflects the nonlinear vorticity dynamics, and it is tempting to study the impact of the coherent vortex structures on this statistical balance.
This has been done with the help of wavelet decomposition techniques in ref.~\cite{wilczek12pof} to which we kindly refer the interested reader.

Also shown in fig.~\ref{fig:1p_vorticity_conave} (right) are the eigenvalues of the conditional enstrophy dissipation tensor, which display a strong $\omega$-dependence.
This is in contrast to the conditional energy dissipation tensor discussed in section~\ref{sec:singlepointvelocity}.
There, the simple scale-separation argument between the velocity and (pseudo-)dissipation fields has turned out to be approximately valid for low velocities.
For the vorticity, much stronger correlations are observed because the vorticity and the gradients of the vorticity vary on comparable scales.
Also the two eigenvalues differ strongly in amplitude (note that $\mu$ and $\lambda/4$ are shown in fig.~\ref{fig:1p_vorticity_conave} (right)).
This is an indication of strong directional correlations of the vorticity and the vorticity gradients related to the filamentary coherent vortex structures.
It is again possible to calculate the stationary solution of~ \eqref{eq:pdfvorticityhomoiso} which takes the form
\begin{equation}\label{eq:statsolvorticity}
  \tilde f(\omega)=\frac{{\cal N}}{\lambda(\omega)} \exp \int_{\omega_0}^\omega \! \mathrm{d}\omega' \, \frac{ -\Sigma(\omega') \, \omega'-\Phi(\omega')+\frac{2}{\omega'}\mu(\omega')}{\lambda(\omega')} \eqspace .
\end{equation}
As we have learned from the conditional balance of vortex stretching and vorticity diffusion, the forcing term $\Phi$ can to good approximation be neglected in this relation.
This is a quantitative proof of the above statement that the single-point vorticity statistics is rather independent of the external forcing mechanism.
As pointed out by Novikov \cite{novikov93jfr}, the influence of the forcing is rather implicit as it maintains the stationarity of the flow.
Novikov, however, also showed that the external forcing is expected to have influence on the two-point vorticity correlations \cite{novikov93prl}.
The validity of relation~\eqref{eq:statsolvorticity} is again tested with our numerical results and, as expected, excellent agreement is demonstrated in fig.~\ref{fig:1p_vorticity_conave}.

To summarize the results of the investigation of the three-dimensional single-point velocity and vorticity statistics, by the joint use of kinetic equations and DNS data it has become possible to account the slight deviations from Gaussianity of the single-point velocity to moderate statistical correlations and the strong non-Gaussianity of the vorticity PDF to more pronounced statistical correlations.
The fact that the vorticity statistics is more independent of the large-scale forcing together with the occurrence of coherent vortex structures indicates that a deeper understanding of the statistics may result from further investigation of vortex dynamics.

\section{Application to Turbulent Rayleigh-B\'{e}nard Convection}
Kinetic equations have been investigated for Rayleigh-B\'{e}nard convection by L{\"u}lff et al.~\cite{luelff11njp}.
Already the single-point kinetic equation yields considerable insight into the connection between statistics and dynamics of turbulent convection, and also allows one to link to phenomenological theories formulated on the basis of dimensional arguments.
Especially, basic assumptions of the Grossmann-Lohse theory \cite{grossmann00jfm} can be investigated by the combination of kinetic equations closed by taking conditional averages from DNS.

\begin{figure}[t]
 \centering{\includegraphics[width=0.7\textwidth]{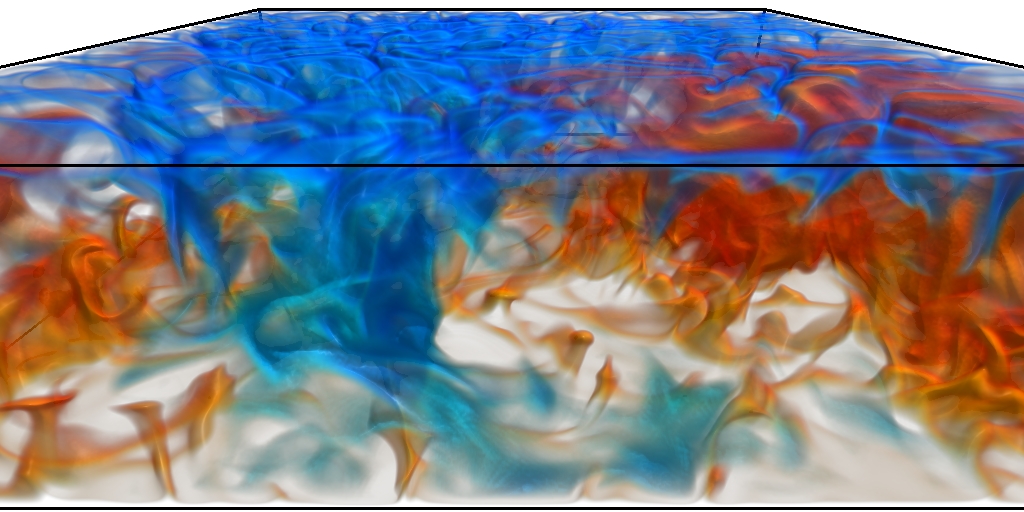}}
 \caption{Visualization of the temperature field in turbulent Rayleigh-B\'{e}nard convection. Red represents hot, blue cold parts of the fluid. Temperatures around the mean are translucent. Parameters are $\Ra=2.4\times10^7$, $\Pr=1$, aspect ratio $\Gamma=4$. Videos of this visualization to be found at \cite{url:turbulenceteamms}.}
 \label{fig:rb_viz}
\end{figure}

Rayleigh-B\'{e}nard convection idealizes the heat transport of a convecting fluid enclosed between a hot bottom and a cold top plate.
It is described by the Oberbeck-Boussinesq equations for the velocity field ${\bm u}({\bm x},t)$ and the temperature field $T({\bm x},t)$:
\begin{align}
  \frac{\partial }{\partial t} T +{\bm u}\cdot \nabla T&=\Delta T \nonumber \\
  \frac{\partial }{\partial t}{\bm u} +{\bm u}\cdot \nabla {\bm u}&=-\nabla p+\Pr\,\Delta {\bm u} + \Pr\,\Ra\, T\, {\bm e}_z
\end{align}
The velocity field is incompressible, $\nabla \cdot {\bm u}=0$, and vanishes at solid boundaries.
Additionally, boundary conditions for the temperature field have to be defined, i.\,e.~a hot bottom and a cold top plate.
Gravity acts in vertical ${\bm e}_z$-direction.
The system is characterized by two dimensionless parameters, the Rayleigh number $\Ra$, which describes the temperature gradient, and the Prandtl number $\Pr$, which describes fluid properties.

To estimate conditional averages from DNS, the full Oberbeck-Boussinesq equations are solved via a pseudospectral method acting on an equidistant, tri-periodic grid.
The boundary conditions in vertical direction, i.\,e.~impermeable plates of constant temperature, are enforced via a suitably designed volume penalization approach (for more information about the method of volume penalization, we refer the reader to e.\,g.~\cite{angot99num,schneider05caf,keetels07jcp}), while in the horizontal directions, periodic boundary conditions are assumed.
A volume rendering of the turbulent temperature field is exhibited in fig.~\ref{fig:rb_viz}, displaying the overall tangled structure of the temperature field, as well as thin filaments of hot and cold fluid emerging from the horizontal plates which are referred to as \emph{plumes}.

\subsection{LMN Hierarchy for Temperature and Velocity}
The full single-point kinetic equation for the joint PDF of temperature and velocity, $f(T, \bm{u},\bm x,t)$, has been derived in \cite{luelff11njp}:
\begin{align}\label{eq:kinetic_eq_rb}
  \frac{\partial}{\partial t} f + {\bm u}\cdot\nabla f = &-\frac{\partial}{\partial T}\Bigl[ \langle\Delta T|T,{\bm u},{\bm x},t \rangle f\Bigr] \nonumber\\
  &-\nabla_{\bm u}\cdot\Bigl[ \bigl( \langle -\nabla p + \Pr\,\Delta {\bm u} | T,{\bm u},{\bm x},t \rangle + \Pr\,\Ra\,T\,{\bm e}_z \bigr)f \Bigr]
\end{align}
For $\Pr=1$, a homogeneity relation similar to \eqref{eq:homrel} can be applied, and the resulting kinetic equation reads
\begin{align}\label{eq:kinetic_eq_homogeniety_relation_rb}
  \frac{\partial }{\partial t}f&+{\bm u}\cdot \nabla f \nonumber\\
  &=\Delta f-\nabla_{\bm u}\cdot\Bigl[\bigl(\langle -\nabla p|\abbrev \rangle + \Ra\,T\,{\bm e}_z \bigr)f\Bigr] - \frac{\partial^2}{\partial T^2} \Bigl[\langle \left(\nabla T\right)^2|\abbrev \rangle f\Bigr] \\ 
  &\hspace*{0.975cm}- 2\frac{\partial^2}{\partial T \partial u_j}\Bigl[ \langle \nabla T\cdot \nabla u_j|\abbrev \rangle f\Bigr] - \frac{\partial^2}{\partial u_i \partial u_j} \Bigl[\langle \nabla u_i\cdot \nabla u_j|\abbrev \rangle f\Bigr] \eqspace,\nonumber
\end{align}
with $\abbrev\,\widehat{=}\,T,{\bm u},{\bm x},t$ abbreviating the dependencies of the conditional averages.
The joint temperature-velocity PDF is effectively determined by the conditionally averaged dissipation-like terms $\langle\left(\nabla T\right)^2|\abbrev \rangle$, $\langle \nabla T\cdot \nabla u_j|\abbrev \rangle$ and $\langle \nabla u_i\cdot \nabla u_i|\abbrev \rangle$, and the conditional pressure gradient $\langle -\nabla p|\abbrev \rangle$.

From the full kinetic equation \eqref{eq:kinetic_eq_rb} one can proceed to the reduced PDF $h(T,z)$ of temperature alone, which after applying the statistical symmetries of homogeneity and stationarity depends on the vertical coordinate $z$ only and is governed by the kinetic equation
\begin{equation}
  \frac{\partial }{\partial z}\Bigl[\langle u_z|T,z\rangle\,h(T,z)\Bigr] + \frac{\partial }{\partial T}\Bigl[ \langle \Delta T|T,z \rangle\,h(T,z) \Bigr] = 0\eqspace.
\end{equation}
Thereby, the average dynamics in $T,z$-phase space and hence the temperature PDF is determined by the conditional averages of vertical velocity $\langle u_z|T,z\rangle$ and heat diffusion $\langle \Delta T|T,z \rangle$.
The associated characteristic equations that depict the aforementioned dynamics read
\begin{subequations}\label{eq:characteritics_rb}
\begin{align}
  \frac{\mathrm{d}}{\mathrm{d}s} T(s) &= \langle \Delta T|T,z \rangle\Bigl.\Bigr\rvert_{\begin{smallmatrix}T=T(s)\\z=z(s)\end{smallmatrix}} \\
  \frac{\mathrm{d}}{\mathrm{d}s} z(s) &= \langle u_z|T,z \rangle\Bigl.\Bigr\rvert_{\begin{smallmatrix}T=T(s)\\z=z(s)\end{smallmatrix}} \\
  \frac{\mathrm{d}}{\mathrm{d}s} h(s) &= -\Bigl(\frac{\partial }{\partial z} \langle u_z|T,z\rangle +\frac{\partial }{\partial T}  \langle \Delta T|T,z \rangle\Bigr)\Bigl.\Bigr\rvert_{\begin{smallmatrix}T=T(s)\\z=z(s)\end{smallmatrix}}h(s)\eqspace ,
\end{align}
\end{subequations}
with $s$ being the parametrization of the characteristic curves; $s$ can be identified with the time $t$ of the system (cf.~section~\ref{sec:method_of_characteristics} and eq.~\eqref{eq:moc_ode_char_t} therein).

\subsection{Numerical Results}
\begin{figure}[t]
 \includegraphics{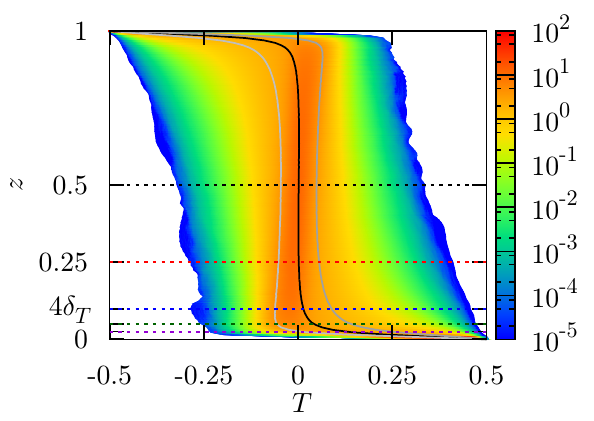}
 \includegraphics{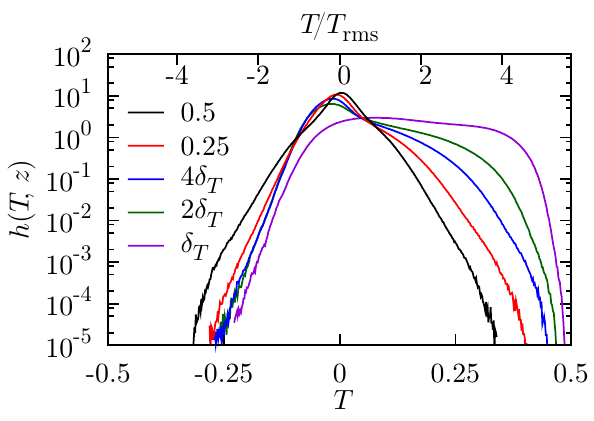}
 \caption{Height-resolved temperature PDF $h(T,z)$ for turbulent Rayleigh-B\'{e}nard convection in logarithmic scaling. Dimensionless parameters as in fig.~\ref{fig:rb_viz}. Left: Two-dimensional color plot. The solid black and gray lines indicate mean and standard deviation of the temperature PDF, respectively. Right: Slices in $T$-direction, as indicated by the horizontal dashed lines in the color plot on the left, with $\delta_T=\frac{1}{2\Nu}$ being the thermal boundary layer thickness and $T_\text{rms}=\sqrt{\langle T^2\rangle_V}$ the quadratic mean of temperature with respect to the whole volume.}
 \label{fig:rb_temp_pdfs}
\end{figure}
In fig.~\ref{fig:rb_temp_pdfs} the $z$-resolved temperature PDF $h(T,z)$ is shown.
In the left panel, the full two-dimensional PDF is color-coded, and one can observe that the PDF shows an almost constant mean value in the bulk regions (i.\,e.~for $4\delta_T\lesssim z \lesssim 1-4\delta_T$) while the skewness varies with $z$ (indeed, it is observed that the skewness varies almost linear in the bulk, cf.~\cite{emran08jfm}).
In the boundary regions the PDF contracts to a $\delta$-function due to the boundary conditions of fixed temperature.
The right panel of fig.~\ref{fig:rb_temp_pdfs} shows one-dimensional cuts at fixed values of $z$ as indicated in the two-dimensional PDF on the left; one can see that the temperature PDF broadens for low values of $z$ before contracting to a $\delta$-function.

\begin{figure}[t]
 \centering{\includegraphics{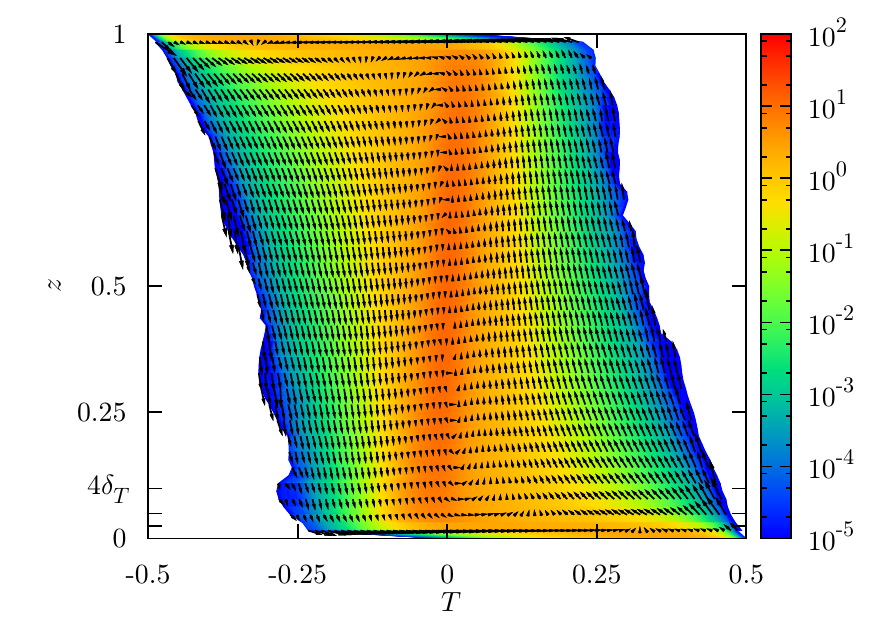}}
 \caption{Vector field $\bigl(T(s),z(s)\bigr)$ governing the characteristic equations in turbulent Rayleigh-B\'{e}nard convection, cf.~eq.~\eqref{eq:characteritics_rb}. Additionally, the temperature PDF is color-coded in compliance with fig.~\ref{fig:rb_temp_pdfs}. Dimensionless parameters as in fig.~\ref{fig:rb_viz}. One can clearly see the generic \emph{RB cycle} discussed in the text.}
 \label{fig:rb_characteristics}
\end{figure}

The vector field that governs the characteristic curves is depicted in fig.~\ref{fig:rb_characteristics}, together with a color plot of the PDF $h(T,z)$.
The characteristic curves show the average behavior of a fluid parcel, i.\,e.~the typical \emph{RB cycle} of fluid heating up at the bottom, rising up to the top plate, cooling down at the top and falling down to the bottom again.
It is tempting to interpret the characteristics as a kind of Lagrangian dynamics of a tracer particle inside the RB cell, in close analogy to the \emph{smart particles} experimentally investigated by the Lyon group (see e.\,g.~\cite{gasteuil07prl}).
However, in contrast to Lagrangian trajectories the characteristics are trajectories in an averaged field and, therefore, only take the chaotic features into account in an averaged sense.

The alternative representation of the heat conductive term similar to the homogeneity relation which leads to \eqref{eq:kinetic_eq_homogeniety_relation_rb} gives rise to the kinetic equation
\begin{equation}
  \frac{\partial }{\partial z}\Bigl[\langle u_z|T,z\rangle\,h(T,z)\Bigr]
  =\frac{\partial^2}{\partial z^2}\,h(T,z) - \frac{\partial^2 }{\partial T^2} \Bigl[\langle \left(\nabla T\right)^2|T,z \rangle\, h(T,z) \Bigr]\eqspace .
\end{equation}
Here, the conditional average $\langle \left(\nabla T\right)^2|T,z \rangle$ comes up.
This term can be related to the Nusselt number, which is defined as the spatial average over the whole cell with volume $V$, $\Nu=\langle (\nabla T)^2 \rangle_V$.
This quantity can be calculated in terms of the conditional average $\langle \left(\nabla T\right)^2|T,z \rangle$ as
\begin{equation}
  \Nu=\frac{1}{V} \int\!\mathrm{d}{\bm x} \int\!\mathrm{d}T \, \langle \left(\nabla T\right)^2|T,z \rangle\,h(T,z)\eqspace .
\end{equation}
Therefore, $\langle \left(\nabla T\right)^2|T,z \rangle$ can be viewed as a conditional Nusselt number.
It is of considerable interest for the evaluation of phenomenological theories concerning the Rayleigh number dependence of the Nusselt number based on a decomposition of the heat transport into bulk and boundary contributions, which underlies the Grossmann-Lohse theory \cite{grossmann00jfm}.
Also, this theory relies on the $z$-resolved dissipation rates, which can be linked to the conditional averages of quadratic terms in \eqref{eq:kinetic_eq_homogeniety_relation_rb} which naturally come up in our derivation.

An approach similar to the one pursued by us has been performed by Yakhot \cite{yakhot89prl} and Ching \cite{ching93prl} for the temperature PDF, and also by Sinai et al.~\cite{sinai89prl} for a passive scalar (in contrast to the temperature as an active scalar in RB convection).
Starting from different premises than we do, Yakhot ends up with a formula that also describes the temperature PDF in terms of conditional averages.
After approximating these averages in the high Rayleigh number limit, he arrives at a temperature PDF of exponential shape in the whole fluid volume so that the PDF does not depend on the vertical coordinate $z$.
In contrast to this result and also to summarize our findings, the temperature PDF that we measured from the numerics clearly shows a $z$-dependence, while also our ansatz is able to connect the statistics of the system, i.\,e.~the different conditional averages that naturally show up in our derivation, to the dynamics of the system, i.\,e.~the characteristic curves that reproduce the distinctive RB cycle.

\section{Burgers Turbulence}
Up to now we have reviewed kinetic equations, which were closed using  conditional expectations taken from DNS.
In this section, we review an application of the LMN hierarchy to the forced Burgers equation, which allows for an exact assessment of the multi-point PDFs, and hence yields considerable insights into the structure of the LMN hierarchy.
Burgers turbulence has always been used as a playground for the evaluation of novel theoretical approaches and concepts to the problem  of turbulence \cite{Bec2007pr}.

The Burgers equation can be exactly solved by application of a Hopf-Cole transformation.
However, the statistical behavior of the forced  Burgers equation is still under heavy debate.
We refer the reader to the review of B\'{e}c and Khanin \cite{Bec2007pr}.
  
Polyakov \cite{Polyakov1995pre} investigated the increment statistics of the driven Burgers equation
\begin{equation}
  \left[\frac{\partial }{\partial t}+u(x,t)\frac{\partial }{\partial x}\right]u(x,t)
  =\nu \frac{\partial^2}{\partial x^2}u(x,t)+F(x,t)
\end{equation}
with a long-range correlated white noise source, based on the cumulative probability $\langle \Theta \left(v-\left [ u(x+r,t)-u(x,t)\right] \right) \rangle$ ($\Theta(v)$ denotes the Heaviside function, $\Theta'(v)=\delta(v)$).
He introduced a closure for the expectation of the conditional dissipation term and was able to suggest a PDF for the velocity gradient.
This work has stimulated subsequent investigations on the treatment of the dissipation anomaly.

In \cite{Eule2006pla} it was possible to find an exact solution of the forced Burgers equation with a stochastic forcing $F(x,t)=\eta(t) x$ that is linear in $x$ and $\eta(t)$ is a Gaussian white noise term with $\langle \eta(t) \eta(t') \rangle = Q \delta(t-t')$.
The solution has the form 
\begin{equation}
  u(x,t)=a(t)x+\Lambda(t) w(\Lambda(t)x,\tau(t)) \eqspace .
\end{equation}
Here, $w(\xi,\tau)$ is an arbitrary solution of the homogeneous Burgers  equation, and the amplitude $a(t)$ obeys the stochastic differential equation $\dot a = - a^2 + \eta$.
The observables $\Lambda(t)$ and $\tau(t)$ are random variables related to the statistics of the amplitude $a(t)$ by $\dot \Lambda=-a\Lambda$ and  $\dot \tau=\Lambda^2$.
The solution exhibits a finite-time singularity, which corresponds to the formation of a new shock.
It is interesting to notice that the $N$-point PDFs can be constructed explicitly in terms of the statistics of $a(t)$ and the initial condition for the solution of the homogeneous Burgers equation $w(\xi,\tau)$.
For the sake of convenience, we formulate the $N$-point PDF as 
\begin{align}
  f(u_1,x_1,&\dots,u_N,x_N,t)=\nonumber \\&\int \! \di a \di \Lambda \di\tau \, h(a,\Lambda,\tau,t) \prod_{i=1}^N \left\langle \delta\bigl(u_i-a x_i -\Lambda w(\Lambda x_i,\tau)\bigr) \right \rangle_w\eqspace,
\end{align}
where $h(a,\Lambda,\tau,t)$ is the solution of the Fokker-Planck equation for the stochastic process
\begin{align}
  \dot{a} (t) & = - a^2 (t) + \eta (t) \nonumber \\
  \Lambda(t) & = e^{-\int_0^t \! \di t' \, a(t')} \nonumber \\
  \tau (t) & =\int_0^t \! \di t''\, e^{-2\int_0^{t''} \! \di t' \, a(t')} \eqspace.
\end{align}
Here, an additional average $\langle \cdot \rangle_w$ with respect to the initial conditions of the solution $w(\xi,\tau)$ has been included.

It is interesting to note that, if we take the initial condition $w=0$, the $N$-point PDF has the form
\begin{align}
  f&(u_1,x_1,\dots , u_N,x_N,t) = \nonumber \\
  & \delta\left(u_1-\frac{x_2}{x_N}u_2\right)\delta\left(u_2-
  \frac{x_3}{x_N}u_3\right) \dots \delta\left(u_{N-1}-\frac{x_{N-1}}{x_N}u_N\right)\frac{1}{x_N} h_0\left(\frac{u_N}{x_N},t\right)\eqspace,
\end{align}
where $h_0$ is the solution of the Fokker-Planck equation
\begin{equation}
  \frac{\partial }{\partial t}h_0(a,t) = \frac{\partial }{\partial a}a^2 h_0(a,t) +
  \frac{Q}{2} \frac{\partial^2}{\partial a^2} h_0(a,t) \eqspace.
\end{equation}
Although the result is rather trivial, due to the trivial behavior of $u(x,t)$, we can conclude that $f(u_1,x_1,\dots , u_N,x_N,t)$ actually describes a Markov chain in scale with a sharp transition probability 
\begin{equation}
  p(u_i,x_i|u_{i+1},x_{i+1})=\delta\left(u_i-\frac{x_{i+1}}{x_N}u_{i+1}\right) \eqspace .
\end{equation}
We remind the reader that such a structure underlies the  phenomenological theory of intermittency formulated by Friedrich and Peinke \cite{Friedrich1997prl} for the direct cascade of turbulence.
It would be interesting to see whether the Markovian property, which can be proven here from first principles, survives for the case, where the forcing $F(x,t)$ has a finite correlation length leading to the emergence of shocks.

\section{Conclusions}
We have presented an overview of recent works on the framework of the Lundgren-Monin-Novikov hierarchy.
After giving a short introduction into the basic methods and concepts, we have reviewed applications to various turbulent systems ranging from the vorticity statistics in two-dimensional turbulence over the single-point velocity and vorticity statistics in three-dimensional turbulence to the temperature statistics in turbulent Rayleigh-B\'{e}nard convection.
In these works, the closure problem has been treated by estimating the unclosed terms from direct numerical simulations.
While this approach does not solve the closure problem, a direct link between basic dynamical features of the system under consideration and the observed statistics is established without further phenomenological assumptions.
By this, insights into various hallmarks of turbulence statistics like the origin of the energy and enstrophy transfer across scales in two-dimensional turbulence, deviations from Gaussianity in three-dimensional turbulence and the height-resolved temperature statistics in turbulent Rayleigh-B\'{e}nard convection have been obtained.
Furthermore, for the linearly random driven Burgers equation we have reviewed an exact solution for the LMN hierarchy.

An interesting point is the fact that the characteristic equations associated to the kinetic equation for the $N$-point Eulerian PDF can be interpreted in a Lagrangian sense as the description of particles moving in an averaged field.
Especially in two-dimensional turbulence, this brings up the notion of screened or quasi-vortices, which contain the statistically averaged structure of the vorticity field.
In the notion of fluid dynamics this directly links the LMN hierarchy to the field of subgrid modeling.
Furthermore, it allows to make contact with Lagrangian models for turbulence like, for example, the tetrad models, reviewed by Pumir and Naso in this volume.

In the future, the investigation of the PDF of velocity increments started by Yakhot \cite{yakhot98pre} should be complemented by an estimation of the arising conditional expectations of pressure gradient and dissipation from DNS.
An assessment of the corresponding kinetic equation would allow to make contact with the phenomenological theories of the turbulent cascade based on the multifractal idea (see for example the review of Chevillard et al.~in this volume) or the phenomenological theory of Friedrich and Peinke \cite{Friedrich1997prl}, based on the stochastic analysis of the statistics of velocity increments.

A shortcoming of the LMN approach clearly originates from the fact that an estimation of conditional PDFs from DNS has up to now only been feasible for one- or two-point PDFs.
It would be interesting to investigate whether a suitable application of more sophisticated methods of data analysis could help to extend the approach to more points leading to a formulation of higher-order kinetic equations.
Another possibility to overcome this restriction is the development of approximate analytical expressions for the conditional expectations.
Up to now, conditional expectations based on Gaussian statistics have been used, which explicitly fail for the two-dimensional inverse cascade as well as the single-point PDF of three-dimensional vorticity.
However, the detailed investigation of the failure sheds considerable light on the necessary extension of the simple conditional Gaussian ansatz.
Both extensions are expected to help to advance significantly our understanding of the statistics of fully developed turbulence.

\section*{Acknowledgments}
This article is dedicated to the memory of Professor Rudolf Friedrich, who passed away unexpectedly on August 16th 2012.
He has been a dedicated teacher, caring mentor and inspiring scientist.
He will be missed by those who knew him.

\bibliographystyle{elsarticle-num}
\bibliography{references}

\end{document}